\documentclass[aps,onecolumn]{revtex4}
\usepackage{amssymb,epsf}
\usepackage{latexsym}

\begin{document}

\title{Thermodynamic instability of topological black holes \\in Gauss-Bonnet
gravity with a generalized electrodynamics}
\author{S. H. Hendi$^{1,2}$\footnote{email address: hendi@shirazu.ac.ir} and S. Panahiyan$^{1}$}
\affiliation{$^1$Physics Department and Biruni Observatory, College of Sciences, Shiraz
University, Shiraz 71454, Iran \\
$^2$Research Institute for Astrophysics and Astronomy of Maragha (RIAAM),
P.O. Box 55134-441, Maragha, Iran }

\begin{abstract}
Motivated by the string corrections on the gravity and
electrodynamics sides, we consider a quadratic Maxwell invariant
term as a correction of the Maxwell Lagrangian to obtain exact
solutions of higher dimensional topological black holes in
Gauss-Bonnet gravity. We first investigate the asymptotically flat
solutions and obtain conserved and thermodynamic quantities which
satisfy the first law of thermodynamics. We also analyze
thermodynamic stability of the solutions by calculating the heat
capacity and the Hessian matrix. Then, we focus on horizon-flat
solutions with adS asymptote and produce a rotating spacetime with
a suitable transformation. In addition, we calculate the conserved
and thermodynamic quantities for asymptotically adS black branes
which satisfy the first law of thermodynamics. Finally, we perform
thermodynamic instability criterion to investigate the effects of
nonlinear electrodynamics in canonical and grand canonical
ensembles.
\end{abstract}

\maketitle

\section{Introduction}\label{Int}

Nonlinear theories have been studied extensively in the context of
various physical phenomena. In other words, because of the
nonlinear nature of most physical systems, linear theories could
not precisely describe the experimental consequences and we must
inevitably consider the nonlinear models.

Classical electrodynamics, whose field equations originate from
the Maxwell Lagrangian, contains various problems which motivate
one to consider nonlinear electrodynamics (NLED). One of the main
problematic items of Maxwell theory is infinite self-energy of a
point-like charge. The first successful Lorentz invariance theory
for solving this problem was introduced by Born and Infeld
\cite{Born}. After Born-Infeld (BI) theory, although some
different NLED have been introduced with various motivations,
their weak field limits lead to Maxwell theory
\cite{BIpapers,PMIpapers,Soleng,HendiJHEP}. Amongst the NLED
theories, the so-called BI-types, whose first nonlinear correction
is a quadratic function of Maxwell invariant, are completely
special \cite{Soleng,HendiJHEP,HendiNLED}. In addition to the
interesting properties of BI-type NLED theories
\cite{BItypeProperties}, it may be shown that these theories may
arise as a low energy limit of heterotic string theory
\cite{BItyprString}, which leads to increased interest. Taking
into account the first leading correction term of BI-type theories
in which originated from the string theory, one can investigate
the effect of NLED versus Maxwell theory
\cite{HendiMomennia,HendiEPJC}.

In this paper, we consider Gauss-Bonnet (GB) gravity as a natural
generalization of Einstein gravity with a quadratic power of
Maxwell invariant in addition to the Maxwell Lagrangian as a
source. On the gravity point of view, string theories in their low
energy limit give rise to effective models of gravity in higher
dimensions which involve higher curvature terms, while from the
electrodynamics viewpoint, a quadratic power of Maxwell invariant
supplements to the Maxwell Lagrangian with the development of
superstring theory. In the weak field limit, GB gravity and the
Lagrangian of the mentioned NLED reduce to Einstein gravity and
Maxwell Lagrangian, respectively.

On the other hand, it has been shown that there is a close
relationship between black hole thermodynamics and the nature of
quantum gravity. For black holes to be physical objects, it is
necessary for them to be stable under external perturbation. There
are two kinds of stability: physical (dynamical) stability and
thermodynamical stability, which is of interest in this paper. The
most interesting parts of black hole thermodynamics are thermal
instability and Hawking--Page phase transition \cite{HPphase}. It
has been shown that there is no Hawking--Page phase transition for
the Schwarzschild--adS black hole whose horizons have vanishing or
negative constant curvature \cite{SchPhase}.

Thermodynamic properties of black holes in GB gravity with NLED
have been studied before \cite{GBNLED}. In this paper, we will
obtain the black hole solutions of GB-Maxwell corrected (GB-MC)
gravity in arbitrary dimensions and discuss thermal instability
conditions.

One may ask the motivation for considering the GB-MC gravity. As
we know, the Maxwell theory has acceptable consequences, to a
large extent, in various physical areas. So, in transition from
the Maxwell theory to NLED, it is allowable to consider the
effects of small nonlinearity variations, not strong effects. In
other words, in order to obtain physical results with experimental
agreements, one should regard the nonlinearity as a correction to
the Maxwell field. Eventually, we should note that, although
various theories of NLED have been created with different
primitive motivations, only their weak nonlinearities have
physical and experimental importance. So the effects of
nonlinearity should be considered as a perturbation to Maxwell
theory. In addition, in a gravitational framework the GB gravity
is a natural generalization of Einstein gravity (not a
perturbation in general) in higher dimensions. One may regard GB
and MC as the corrections of an Einstein-Maxwell black hole.

The layout of the paper will be in this order: In Sec. \ref{Top},
we will introduce the structure of action which contains GB
gravity coupled with MC. Then, we will solve the field equations
and discuss the geometric properties of the topological black
holes. We will study thermodynamic properties and stability of the
asymptotically flat spacetime in Sec. \ref{Asympflat}. In the next
section, we will generalize our solutions to rotating black branes
with adS asymptote and calculate thermodynamic and conserved
quantities of rotating solutions. Last part is devoted to
investigating the stability of the solutions in canonical and
grand canonical ensembles. Conclusions are drawn in the last
section.

\section{Topological Black Holes} \label{Top}

The Lagrangian of Einstein-GB gravity coupled to NLED can be
written as
\begin{equation}
L_{\mathrm{tot}}=L_{E}-2\Lambda +\alpha L_{GB}+L(F),  \label{Lagrangian}
\end{equation}%
where the Lagrangian of Einstein gravity is the Ricci scalar, $L_{E}=R$, and
$\Lambda $ is the cosmological constant. In the third term of Eq. (\ref%
{Lagrangian}), $\alpha $ is the GB coefficient with dimension
(length)$^{2}$ and $L_{GB}$ is the Lagrangian of GB gravity,
\begin{equation}
L_{GB}=R_{abcd}R^{abcd}-4R_{ab}R^{ab}+R^{2}.  \label{LGB}
\end{equation}

The last term in Eq. (\ref{Lagrangian}) is the Lagrangian of NLED,
which we choose a quadratic correction in addition to the Maxwell
Lagrangian
\begin{equation}
L(\mathcal{F})=-\mathcal{F}+\beta \mathcal{F}^{2}+O(\beta ^{2}),
\label{L(F)}
\end{equation}
where $\beta $ is the nonlinearity parameter and the Maxwell invariant $\mathcal{F%
}=F_{ab}F^{ab}$, where $F_{ab}=\partial _{a}A_{b}-\partial
_{b}A_{a}$ is the electromagnetic field tensor and $A_{b}$ is the
gauge potential. For the vanishing the nonlinearity parameter,
this Lagrangian yields the standard Maxwell theory, as it should.
Regarding the gauge-gravity Lagrangian (\ref{Lagrangian}) and
using the variational method, one can obtain the following field
equations:
\begin{equation}
G_{ab}^{E}+\Lambda g_{ab}+\alpha G_{ab}^{GB}=\frac{1}{2}g_{ab}L(\mathcal{F}%
)-2L_{\mathcal{F}}F_{ac}F_{b}^{c},  \label{Feq1}
\end{equation}%
\begin{equation}
\partial _{a}\left( \sqrt{-g}L_{\mathcal{F}}F^{ab}\right) =0,  \label{Feq2}
\end{equation}%
where $G_{ab}^{E}$ is the Einstein tensor, $G_{ab}^{GB}=2\left(
R_{acde}R_{b}^{cde}-2R_{acbd}R^{cd}-2R_{ac}R_{b}^{c}+RR_{ab}\right) -\frac{1%
}{2}L_{GB}g_{ab}$ and $L_{\mathcal{F}}=\frac{dL(\mathcal{F})}{d\mathcal{F}}$.

Now, we are interested in topological static black hole solutions;
therefore, we take into account the following static metric:
\begin{equation}
ds^{2}=-f(r)dt^{2}+\frac{dr^{2}}{f(r)}+r^{2}d\Omega _{k}^{2},  \label{Metric}
\end{equation}
where $d\Omega _{k}^{2}$ represents the line element of an $(n-1)$%
-dimensional hypersurface with the constant curvature
$(n-1)(n-2)k$ and volume $V_{n-1}$ with the following explicit
form:
\begin{equation}
d\Omega _{k}^{2}=\left\{
\begin{array}{cc}
d\theta _{1}^{2}+\sum\limits_{i=2}^{n-1}\prod\limits_{j=1}^{i-1}\sin
^{2}\theta _{j}d\theta _{i}^{2} & k=1 \\
d\theta _{1}^{2}+\sinh ^{2}\theta _{1}d\theta _{2}^{2}+\sinh ^{2}\theta
_{1}\sum\limits_{i=3}^{n-1}\prod\limits_{j=2}^{i-1}\sin ^{2}\theta
_{j}d\theta _{i}^{2} & k=-1 \\
\sum\limits_{i=1}^{n-1}d\phi _{i}^{2} & k=0%
\end{array}%
\right. .  \label{dOmega}
\end{equation}

Since we are looking for the black hole solution with a radial
electric field, we know that the nonzero components of the
electromagnetic field are
\begin{equation}
F_{tr}=-F_{rt}.  \label{nonzero}
\end{equation}

One can use Eq. (\ref{Feq2}) to obtain the explicit form of
$F_{tr}$ with the following form:
\begin{equation}
F_{tr}=\frac{q}{r^{n-3}}-\frac{4q^{3}\beta }{r^{3n-3}}+O(\beta
^{2}), \label{Ftr}
\end{equation}%
which for small values of $\beta$ reduces to Maxwell
electromagnetic field tensor.

We find that the nonzero independent components of Eq.
(\ref{Feq1}) can be written by applying electromagnetic field
tensor (\ref{Ftr}):
\begin{eqnarray}
e_{tt} &=&r^{3}\left( 1+\frac{2(1-f)\alpha ^{\prime }}{r^{2}}\right)
f^{\prime }-\alpha ^{\prime }(n-4)(1-f)^{2}+  \nonumber \\
&&\frac{2\Lambda r^{4}}{(n-1)}+\frac{2q^{2}}{(n-1)r^{2n-6}}-\frac{%
4q^{4}\beta }{(n-1)r^{4n-8}}+O(\beta ^{2}).  \label{ett}
\end{eqnarray}%
\begin{eqnarray}
e_{\theta \theta } &=&r^{4}\left[ 1+\frac{2(1-f)\alpha ^{\prime }}{r^{2}}%
\right] f^{\prime \prime }-2\alpha ^{\prime }r^{2}f^{\prime 2}  \nonumber \\
&&+2r(n-2)\left[ r^{2}+2\alpha ^{\prime }\frac{(n-4)}{(n-2)}(1-f)\right]
f^{\prime }  \nonumber \\
&&-\alpha ^{\prime }(n-4)(n-5)(1-f)^{2}-(n-2)(n-3)r^{2}(1-f)  \nonumber \\
&&+2\Lambda r^{4}-\frac{2q^{2}}{r^{2n-6}}+\frac{12q^{4}\beta }{r^{4n-8}}%
+O(\beta ^{2}).  \label{ethth}
\end{eqnarray}%
where $\alpha ^{\prime }=(n-2)(n-3)\alpha $

It is straightforward to show that the following metric function
satisfies all of the field equations simultaneously:
\begin{equation}
f(r)=k+\frac{r^{2}}{2\alpha ^{\prime }}\left( 1-\sqrt{\Psi (r)}\right) ,
\label{f(r)}
\end{equation}%
with
\begin{equation}
\Psi (r)=1+\frac{8\alpha ^{\prime }}{n(n-1)}\left( \Lambda +\frac{n(n-1)m}{%
2r^{n}}-\frac{nq^{2}}{(n-2)r^{2n-2}}+\frac{2nq^{4}\beta }{r^{4n-4}(3n-4)}%
\right) +O\left( \beta ^{2}\right) ,  \label{Psi(r)}
\end{equation}%
where $m$ is an integration constant that is related to mass. One
can see that for small values of $\beta$, the metric function
reduces to usual GB-Maxwell gravity. Expansion of the metric
function for small values of the GB parameter leads to
\begin{equation}
f\left( r\right) =f_{EM}-\frac{4q^{4}}{\left( n-1\right) \left( 3n-4\right)
r^{4n-6}}\beta +\frac{(k-f_{EM})^{2}}{r^{2}}\alpha ^{\prime }+\frac{%
8q^{4}(k-f_{EM})}{\left( n-1\right) \left( 3n-4\right) r^{4n-4}
}\alpha ^{\prime }\beta +O\left( \alpha ^{\prime 2},\beta
^{2}\right) , \label{f(r)expand}
\end{equation}%
where the metric function of Einstein--Maxwell gravity is
\begin{equation}
f_{EM}=k-\frac{2\Lambda r^{2}}{n\left( n-1\right) }-\frac{m}{r^{n-2}}+\frac{%
2q^{2}}{\left( n-1\right) \left( n-2\right) r^{2n-4}}.  \label{f(r)EM}
\end{equation}

In order to consider the asymptotic behavior of the solution, we put $m=q=0$
where the metric function reduces to
\begin{equation}
f(r)=k+\frac{r^{2}}{2\alpha ^{\prime }}\left( 1-\sqrt{1+\frac{8\alpha
^{\prime }\Lambda }{n(n-1)}}\right) .  \label{f(r)asymp}
\end{equation}

This result puts a restriction on $\alpha^{\prime }$ which states that for $%
\alpha ^{\prime }\leq \frac{-n(n-1)}{8\Lambda}$, the metric
function will be real asymptotically. On the other hand, for
$\Lambda =0$, asymptotically flat solutions are available only for
$k=1$.

The next step is to look for the singularities. It is a matter of
calculation to show that the Kretschmann scalar of metric
(\ref{Metric}) is
\begin{equation}
R_{\alpha \beta \gamma \delta }R^{\alpha \beta \gamma \delta }=f^{\prime
\prime 2}+2(n-1)\frac{f^{\prime 2}}{r^{2}}+2(n-1)(n-2)\frac{f^{2}}{r^{4}}%
\text{,}  \label{RR}
\end{equation}%
where its series expansion for small and large values of $r$ will be
\begin{equation}
{\lim_{r\longrightarrow 0}}R_{\alpha \beta \gamma \delta }R^{\alpha \beta
\gamma \delta }\propto r^{-4\left( n-1\right) }  \label{RRzero}
\end{equation}%
\begin{equation}
{\lim_{r\longrightarrow \infty }}R_{\alpha \beta \gamma \delta }R^{\alpha
\beta \gamma \delta }=\frac{4(n+2)\Lambda }{n\alpha ^{\prime }}-\frac{%
\varsigma }{\alpha ^{\prime }}\left( 1+\frac{\left( n-1\right) (n+2)}{\alpha
^{\prime }}\right) ,  \label{RRinfinity}
\end{equation}%
\begin{equation}
\varsigma =\sqrt{\frac{n-2}{n}+\frac{8\alpha ^{\prime }\Lambda }{n(n-1)}}-1,
\end{equation}%
which, for small values of the GB parameter, it will lead to
\begin{equation}
{\lim_{r\longrightarrow \infty }}R_{\alpha \beta \gamma \delta }R^{\alpha
\beta \gamma \delta }=\frac{8(n+2)}{(n+2)n^{2}}\Lambda ^{2}-\frac{4}{n\left(
n-1\right) }\Lambda +O\left( \alpha ^{\prime }\right)  \label{RRexpand}
\end{equation}

Eq. (\ref{RRzero})-(\ref{RRexpand}) show that there is an
essential singularity located at $r=0$ and the asymptotical
behavior of the solutions are adS with an effective cosmological
constant. If this solution contains the horizon then our metric
function is interpreted as a black hole. In order to
investigate the existence of the horizon, one should find the root of $%
g^{rr}=f(r)=0.$ If we suppose that the metric function is positive
for large and small $r$, by considering the possibility of the
existence of only one extreme root $r_{+}=r_{ext}$, we know that
$f(r)$ has a minimum at $r=r_{ext}$. So, we can investigate the
roots of $f^{\prime }$,
\begin{equation}
k(n-1)(n-2)r_{ext}^{4n-6}\left[ \frac{\alpha ^{\prime }(n-4)}{%
(n-2)r_{ext}^{2}}+1\right] +4q^{4}\beta
-2r_{ext}^{2n-2}q^{2}-2r_{ext}^{4n-4}\Lambda +O\left( \beta
^{2}\right) =0, \label{rextEq}
\end{equation}%
where, for example, in the Ricci-flat case ($k=0$), we obtain
\begin{equation}
r_{ext}=\frac{\left[ q^{2}\Lambda ^{2n-3}(\pm \sqrt{1+8\Lambda
\beta }-1)\right]^{\frac{1}{2n-2}}}{\Lambda }+O\left( \beta
^{2}\right) .  \label{rextk0}
\end{equation}

It is matter of calculation to show that the extremal mass for
$k=0$ is
\begin{equation}
m_{ext}=\frac{2q^{2}}{r_{ext}^{n-2}(n-1)(n-2)}-\frac{2r_{ext}^{n}\Lambda }{%
n(n-1)}-\frac{4q^{4}\beta }{r_{ext}^{3n-4}(3n-4)(n-1)}+O\left(
\beta ^{2}\right) .  \label{mextk0}
\end{equation}

We should note that the obtained solutions may be interpreted as
black holes with inner and outer horizons provided $m>m_{ext}$, an
extreme black hole for $m=m_{ext}$, and naked singularity
otherwise. One can obtain the Hawking temperature by surface
gravity definition with the following explicit form:
\begin{equation}
T=\frac{k(n-1)(n-2)r_{+}^{4n-6}\left( 1+\frac{(n-4)\alpha ^{\prime }}{%
(n-2)r_{+}^{2}}\right) -2r_{+}^{4n-4}\Lambda
-2r_{+}^{2n-2}q^{2}+4q^{4}\beta }{4\pi (n-1)r_{+}^{4n-5}\left( 1+\frac{%
2k\alpha ^{\prime }}{r_{+}^{2}}\right) }+O\left( \beta ^{2}\right)
. \label{T}
\end{equation}%
It is worthwhile to mention that extremal black holes have zero temperature.

\section{Thermodynamics of Asymptotically Flat Black Holes ($k=1$ \& $\Lambda
=0$)} \label{Asympflat}

In this section we are interested in the thermodynamics of
asymptotically flat solutions. Using the series expansion of the
metric function (\ref{f(r)}) with $k=1$ for large values of
distance, we obtain
\begin{equation}
f(r)=1-\frac{m}{r^{n-2}}+\frac{2q^{2}}{(n-1)(n-2)r^{2n-4}}+O\left( \frac{1}{%
r^{2n-2}}\right) ,  \label{f(r)flat}
\end{equation}%
which shows that the solutions are asymptotically flat.

According to area law, the entropy of black holes is one-quarter
of the horizon area. This relation is acceptable for Einstein
gravity, whereas we are not allowed to use it for higher
derivative gravity. For our GB case we can use the Wald formula
for calculating the entropy of asymptotically flat black hole
solutions
\begin{equation}
S=\frac{1}{4}\int d^{n-1}x\sqrt{\gamma }\left( 1+2\alpha \widetilde{R}\right)
\label{Swald1}
\end{equation}%
where $\widetilde{R}$ is the Ricci scalar for the induced metric
$\gamma _{ab}$ on the $\left( n-1\right) $ dimensional boundary.
Calculations show that one can obtain

\begin{equation}
S=\frac{V_{n-1}}{4}\left( 1+\frac{2\left( n-1\right) \alpha ^{\prime }}{%
(n-3)r_{+}^{2}}\right) r_{+}^{n-1},  \label{Swald2}
\end{equation}%
which confirms that asymptotically flat black holes violate the area law.

Considering the flux of the electric field at infinity, one can
find the electric charge of black holes with the following form:
\begin{equation}
Q=\frac{q}{4\pi }.  \label{Qk1}
\end{equation}

Next, for the electric potential, $\Phi ,$ we use the following definition
\begin{equation}
\Phi =\left. A_{\mu }\chi ^{\mu }\right\vert _{r\longrightarrow \infty
}-\left. A_{\mu }\chi ^{\mu }\right\vert _{r\longrightarrow r_{+}}=\frac{q}{%
(n-2)r_{+}^{n-2}}\left( 1-\frac{4\left( n-2\right) q^{2}\beta
}{\left( 3n-4\right) r_{+}^{2n-2}}\right) +O\left( \beta
^{2}\right) , \label{Phik1}
\end{equation}%
where $\chi ^{\mu }$ is the null generator of the horizon. It is
notable that although the electric potential depends on the
nonlinearity of electrodynamics, the electric charge does not.

In order to calculate the finite mass of the black hole we use the
ADM (Arnowitt-Deser-Misner) approach for large values of $r$,
which will result in \cite{Brewin}
\begin{equation}
M=\frac{V_{n-1}}{16\pi }m\left( n-1\right) .  \label{Massk1}
\end{equation}
We should note that one can obtain $m$ from $f(r=r_{+})=0$, so the
total mass depends on GB and NLED parameters.

Next, by using Eqs. (\ref{f(r)}), (\ref{Swald2}), (\ref{Qk1}) and
considering $M$ as a function of the extensive parameters $S$ and
$Q$, we have
\begin{equation}
M\left( S,Q\right) =\frac{r_{+}^{n-2}}{16\pi }\left[ (n-1)\left( 1+\frac{%
\alpha ^{\prime }}{r_{+}^{2}}\right) +\frac{32\pi ^{2}Q^{2}}{%
(n-2)r_{+}^{2n-4}}-\frac{1024\pi ^{4}Q^{4}\beta }{(3n-4)r_{+}^{4n-6}}%
\right] +O\left( \beta ^{2}\right) .  \label{Msmar}
\end{equation}

Now, we calculate the temperature and electric potential as the intensive
parameter with the following relation
\begin{equation}
T=\left( \frac{\partial M}{\partial S}\right) _{Q}=\left( \frac{\partial M}{%
\partial r_{+}}\right) _{Q}\left/ \left( \frac{\partial S}{\partial r_{+}}%
\right) _{Q}\right. \ \ ,\ \ \Phi =\left( \frac{\partial M}{\partial Q}%
\right) _{S},\   \label{TPhi}
\end{equation}%
which are the same as the ones that were calculated in Eqs. (\ref{T}) and (\ref%
{Phik1}). Thus, the conserved and thermodynamic quantities satisfy the first
law of thermodynamics,
\begin{equation}
dM=TdS+\Phi dQ.  \label{Firstk1}
\end{equation}

\subsection{Stability of the solutions}

Here, we investigate the thermodynamic stability of charged black
hole solutions of GB gravity with NLED. In the grand canonical
ensemble, the thermodynamic stability may be carried out by
calculating the determinant of
the Hessian matrix of $M(S,Q)$ with respect to its extensive variables $%
X_{i} $, $\mathbf{H}_{X_{i}X_{j}}^{M}$ $=\left( \frac{\partial ^{2}M}{%
\partial X_{i}\partial X_{j}}\right) $. In the canonical ensemble, the
electric charge is a fixed parameter, and therefore the positivity
of the heat capacity $C_{Q}=T_{+}/\left( \frac{\partial
^{2}M}{\partial S^{2}}\right) $ is sufficient to ensure the local
stability. Since the physical black hole solutions have positive
temperature, it is sufficient to check the positivity of $\left(
\frac{\partial ^{2}M}{\partial S^{2}}\right)
_{Q}=\left( \frac{\partial T}{\partial r_{+}}\right) _{Q}/\left( \frac{%
\partial S}{\partial r_{+}}\right) _{Q}$,
\begin{eqnarray}
\left( \frac{\partial ^{2}M}{\partial S^{2}}\right) _{Q} &=&-\left\{
r_{+}^{4n-8}(n-1)\left( n-2\right) \left[ \frac{2(n-4)\alpha ^{\prime 2}}{%
(n-2)}+\frac{(n-8)\alpha ^{\prime }r_{+}^{2}}{(n-2)}+r_{+}^{4}\right] \right.
\nonumber \\
&&\left. -2r_{+}^{2n-2}q^{2}\left[ 2\alpha ^{\prime }(2n-5)+(2n-3)r_{+}^{2}%
\right] +2\beta q^{4}\left[ 4\alpha ^{\prime }\right. \right.  \nonumber \\
&&\left. \left. (4n-7)+(8n-10)r_{+}^{2}\right] \right\} /\left[ \pi \left(
n-1\right) ^{2}r_{+}^{5n-10}\left( 2\alpha ^{\prime }+r_{+}^{2}\right) ^{3}%
\right] +O\left( \beta ^{2}\right) .  \label{dMdSSk1}
\end{eqnarray}

One may study the behavior of $\left( \frac{\partial
^{2}M}{\partial S^{2}}\right)_{Q}$ for small values of the GB
parameter with the following form:
\begin{eqnarray}
\left( \frac{\partial ^{2}M}{\partial S^{2}}\right) _{Q} &=&-\frac{%
(n-1)(n-2)r_{+}^{4n-4}-2(2n-3)r_{+}^{2n}q^{2}}{\left( n-1\right)
^{2}r_{+}^{5n-4}\pi }-\frac{4\left( 4n-5\right) q^{4}}{\left( n-1\right)
^{2}r_{+}^{5n-6}\pi }\beta  \nonumber \\
&&+\frac{\left[ \left( 5n-4\right) r_{+}^{4n}-16r_{+}^{2n+4}q^{2}\right] }{%
\left( n-1\right) r_{+}^{5n+2}\pi }\alpha ^{\prime }+\frac{64q^{4}}{\left(
n-1\right) r_{+}^{5n-4}\pi }\beta \alpha ^{\prime }  \nonumber \\
&&+O\left( \alpha ^{\prime 2},\beta ^{2}\right) .
\label{dMdSSk1bast}
\end{eqnarray}

In addition, it is worthwhile to mention that for small and large
values of $r_{+}$, we obtain
\begin{eqnarray}
\left. \left( \frac{\partial ^{2}M}{\partial S^{2}}\right) _{Q}\right\vert _{%
\text{Small }r_{+}} &=&\left\{
\begin{array}{cc}
-\frac{(4n-7)\beta q^{4}}{\pi (n-1)^{2}\alpha ^{\prime
2}r_{+}^{5n-10}}<0,
& \alpha ^{\prime }\neq 0,\beta \neq 0 \\
-\frac{4\left( 4n-5\right) \beta q^{4}}{\pi \left( n-1\right)
^{2}r_{+}^{5n-6}}<0, & \alpha ^{\prime }=0,\beta \neq 0 \\
\frac{(2n-5)q^{2}}{2\pi \left( n-1\right) ^{2}\alpha ^{\prime
2}r_{+}^{3n-8}}>0, & \alpha ^{\prime }\neq 0,\beta =0 \\
\frac{2(2n-3)q^{2}}{\pi \left( n-1\right) ^{2}r_{+}^{3n-4}}>0, &
\alpha ^{\prime }=0,\beta =0
\end{array}
\right. ,  \label{lower1} \\
\left. \left( \frac{\partial ^{2}M}{\partial S^{2}}\right)
_{Q}\right\vert _{\text{Large }r_{+}} &=&-\frac{n-2}{\pi
(n-1)r_{+}^{n}}<0,  \label{upper1}
\end{eqnarray}
which indicate that for nonzero $\alpha ^{\prime }$ and $\beta$,
asymptotically flat black holes with a large or small horizon
radius are not stable. In other words, there is an upper limit,
$r_{+\max }$, as well as a lower limit, $r_{+\min}$, for the
asymptotically flat stable black holes
($r_{+\min}<r_{+}<r_{+\max}$). Although this result does not
depend on the value of the GB parameter, asymptotically flat black
holes with a small horizon radius may be stable in the presence of
a pure Maxwell field (absence of nonlinearity of electrodynamic
correction). In other words, taking into account thermal stability
in the canonical ensemble, the nonlinearity of electrodynamics has
a reasonable influence on a small horizon radius.

\begin{figure}[tbp]
$
\begin{array}{cc}
\epsfxsize=7.5cm \epsffile{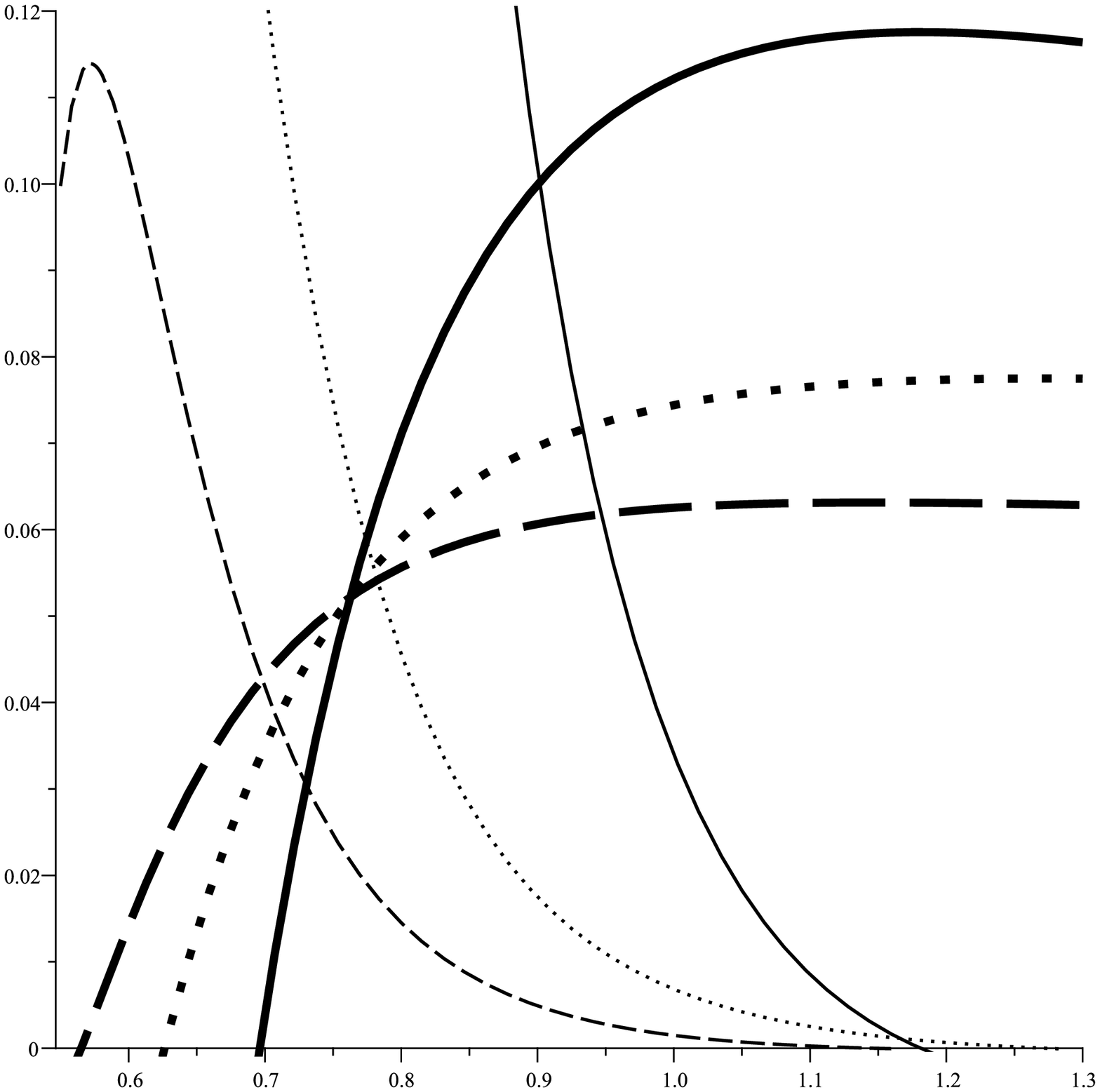} & \epsfxsize=7.5cm %
\epsffile{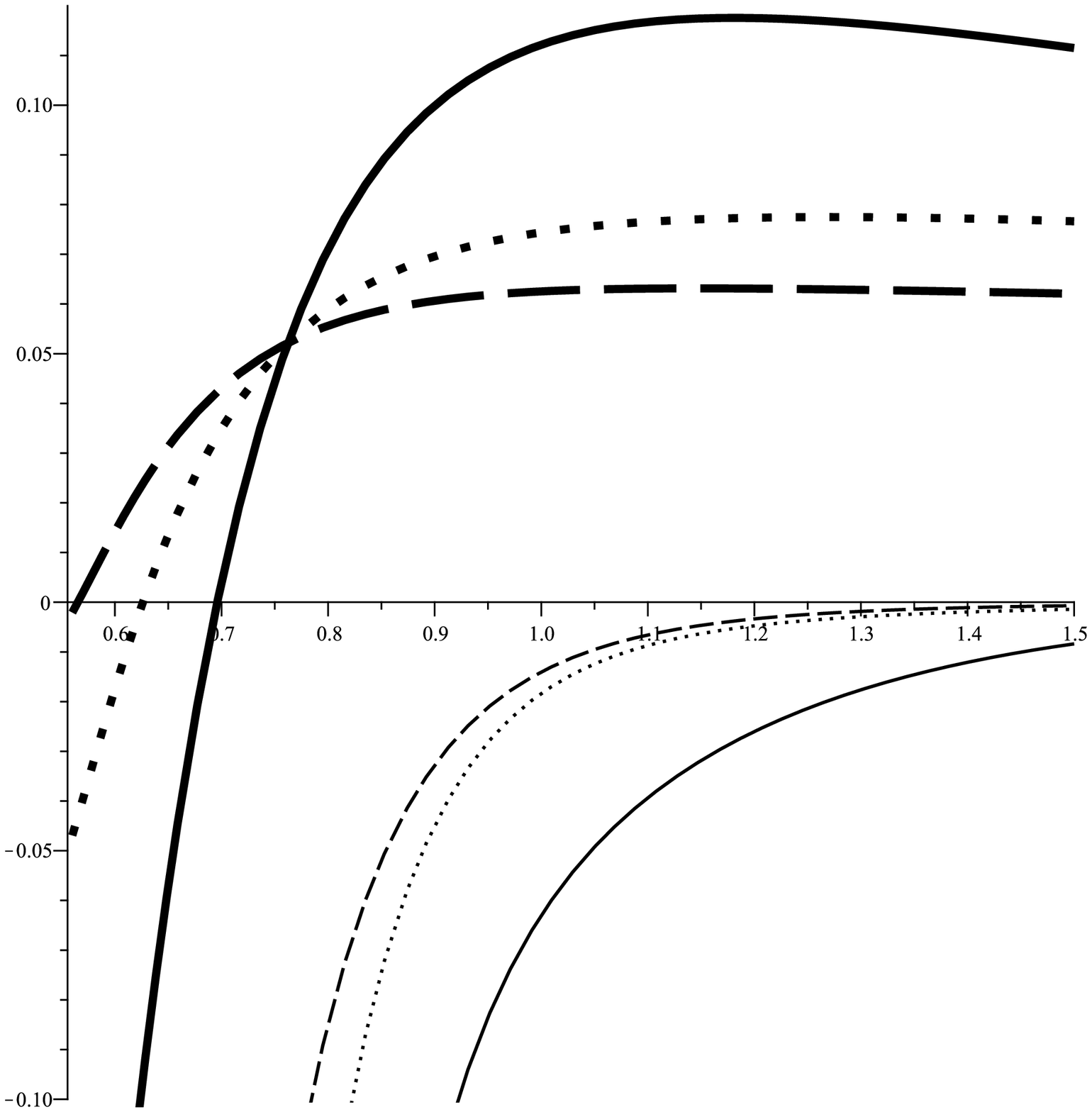}%
\end{array}
$%
\caption{\emph{Asymptotically flat solutions:} $\left( \frac{\partial ^{2}M}{%
\partial S^{2}}\right) _{Q}$ (left figure) and $\left\vert \mathbf{H}%
_{S,Q}^{M}\right\vert$ (right figure) versus $r_{+}$ for $n=5$, $q=1$, $%
\protect\beta=0.001$ and $\protect\alpha=0.1$ (solid line), $\protect\alpha%
=0.3$ (dotted line) and $\protect\alpha=0.5$ (dashed line).
\textbf{"bold lines represent the temperature"}}
\label{Stabilityalpha1}
\end{figure}

\begin{figure}[tbp]
$%
\begin{array}{cc}
\epsfxsize=7.5cm \epsffile{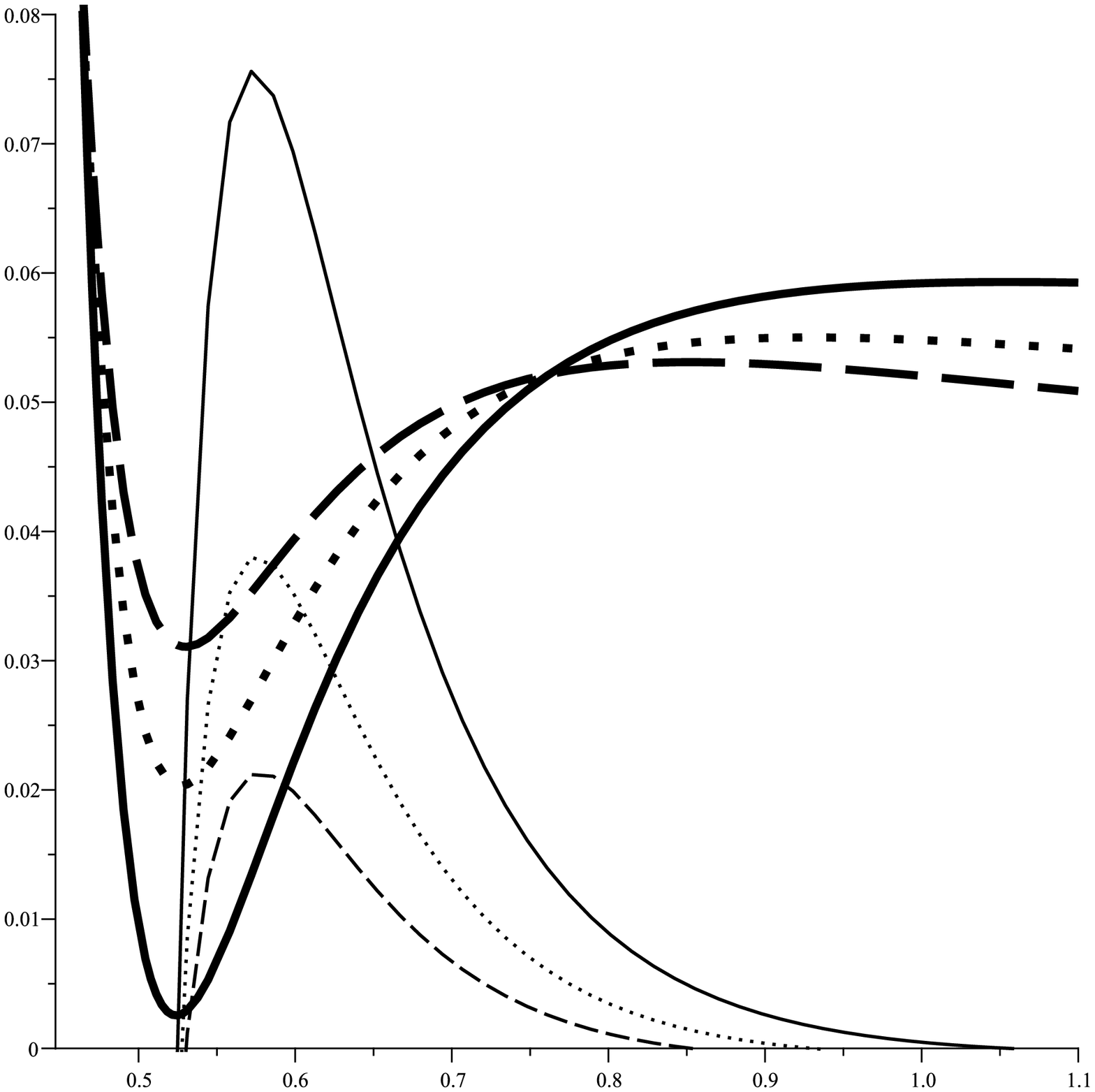} & \epsfxsize=7.5cm %
\epsffile{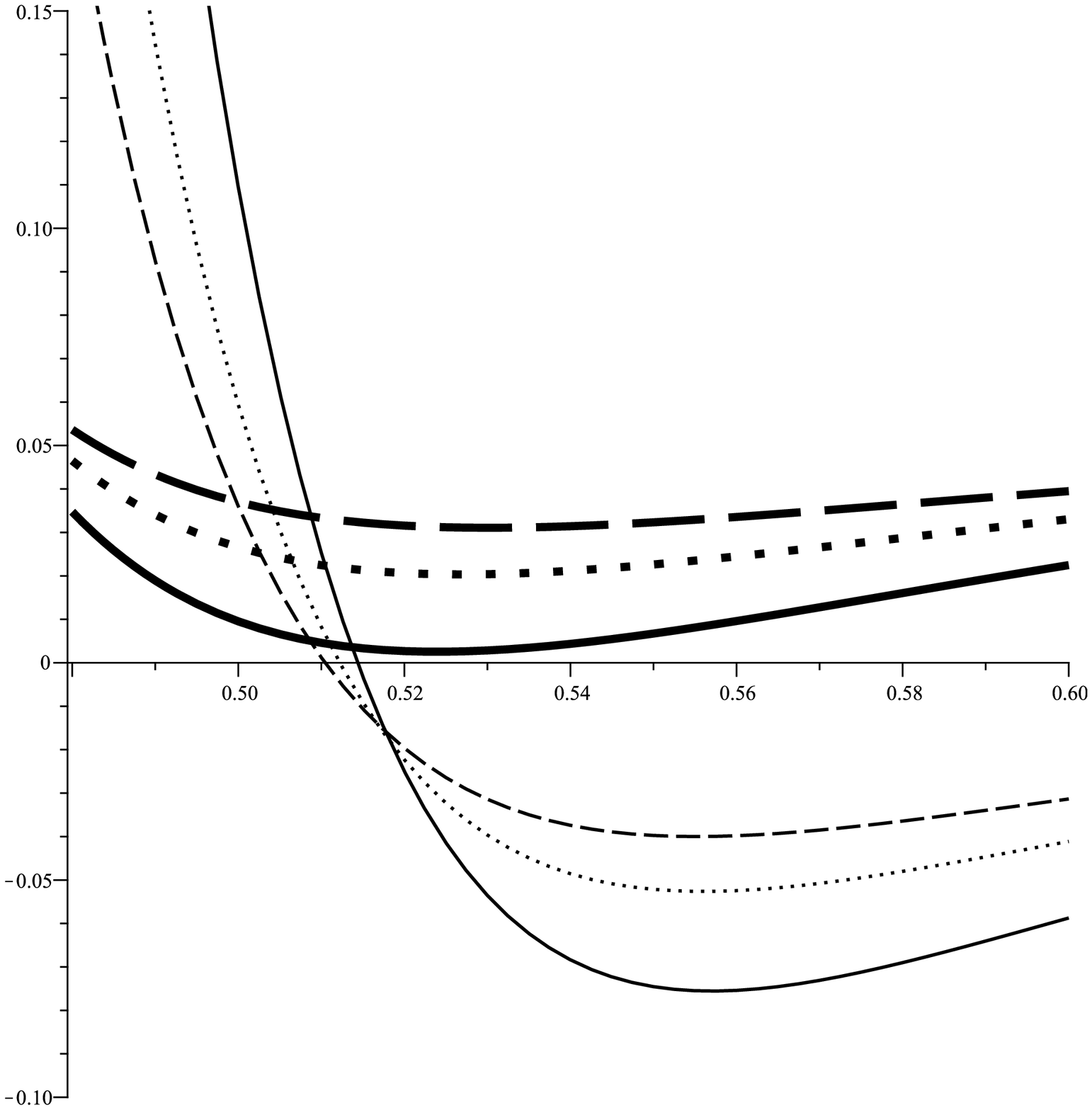}%
\end{array}
$%
\caption{\emph{Asymptotically flat solutions:} $\left( \frac{\partial ^{2}M}{%
\partial S^{2}}\right) _{Q}$ (left figure) and $\frac{\left\vert \mathbf{H}%
_{S,Q}^{M}\right\vert}{10}$ (right figure) versus $r_{+}$ for $n=5$, $q=1$, $%
\protect\beta=0.001$ and $\protect\alpha=0.6$ (solid line), $\protect\alpha%
=0.8$ (dotted line) and $\protect\alpha=0.1$ (dashed line).
\textbf{"bold lines represent the temperature"}}
\label{Stabilityalpha2}
\end{figure}

\begin{figure}[tbp]
$%
\begin{array}{cc}
\epsfxsize=7.5cm \epsffile{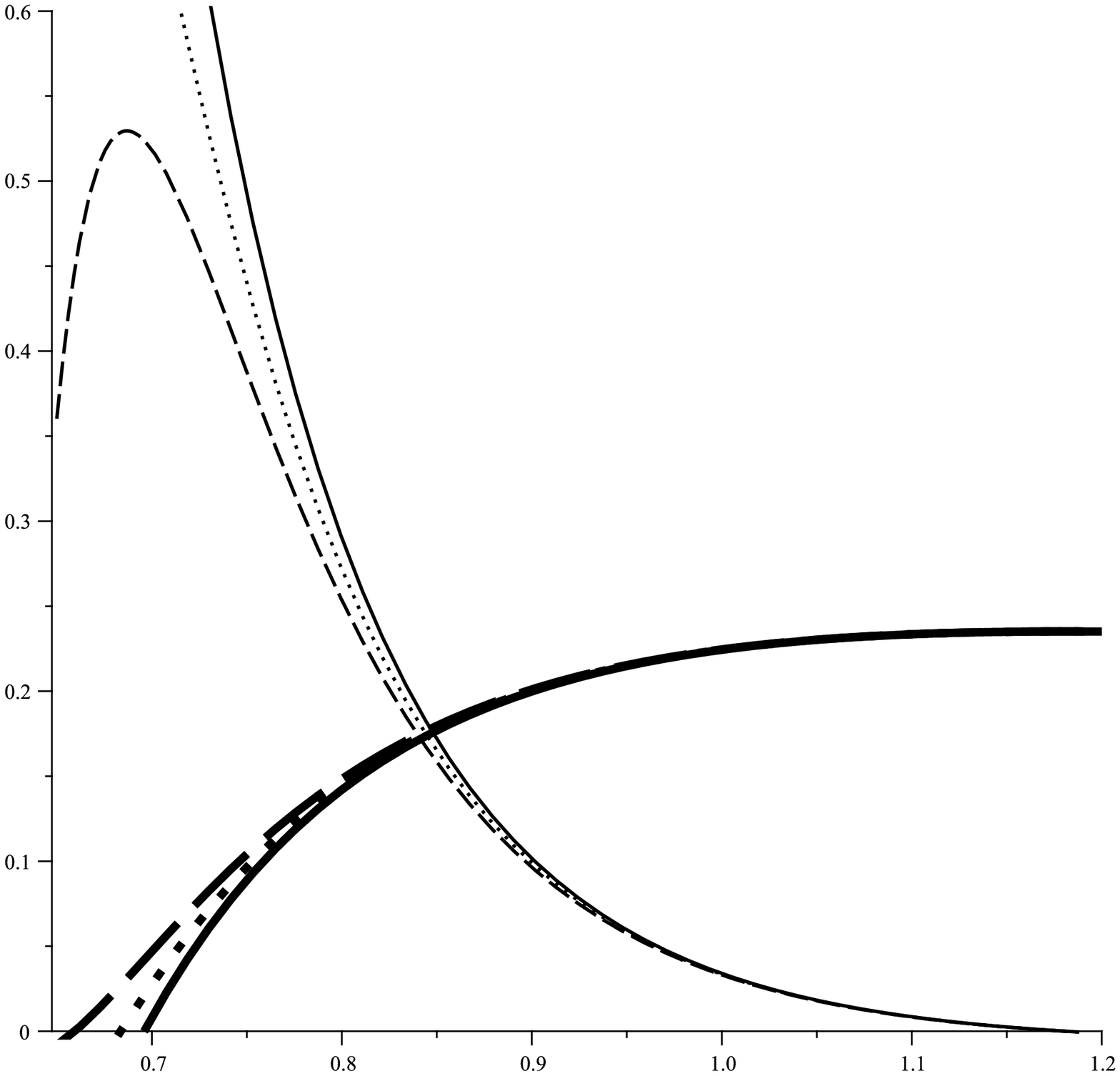} & \epsfxsize=7.5cm %
\epsffile{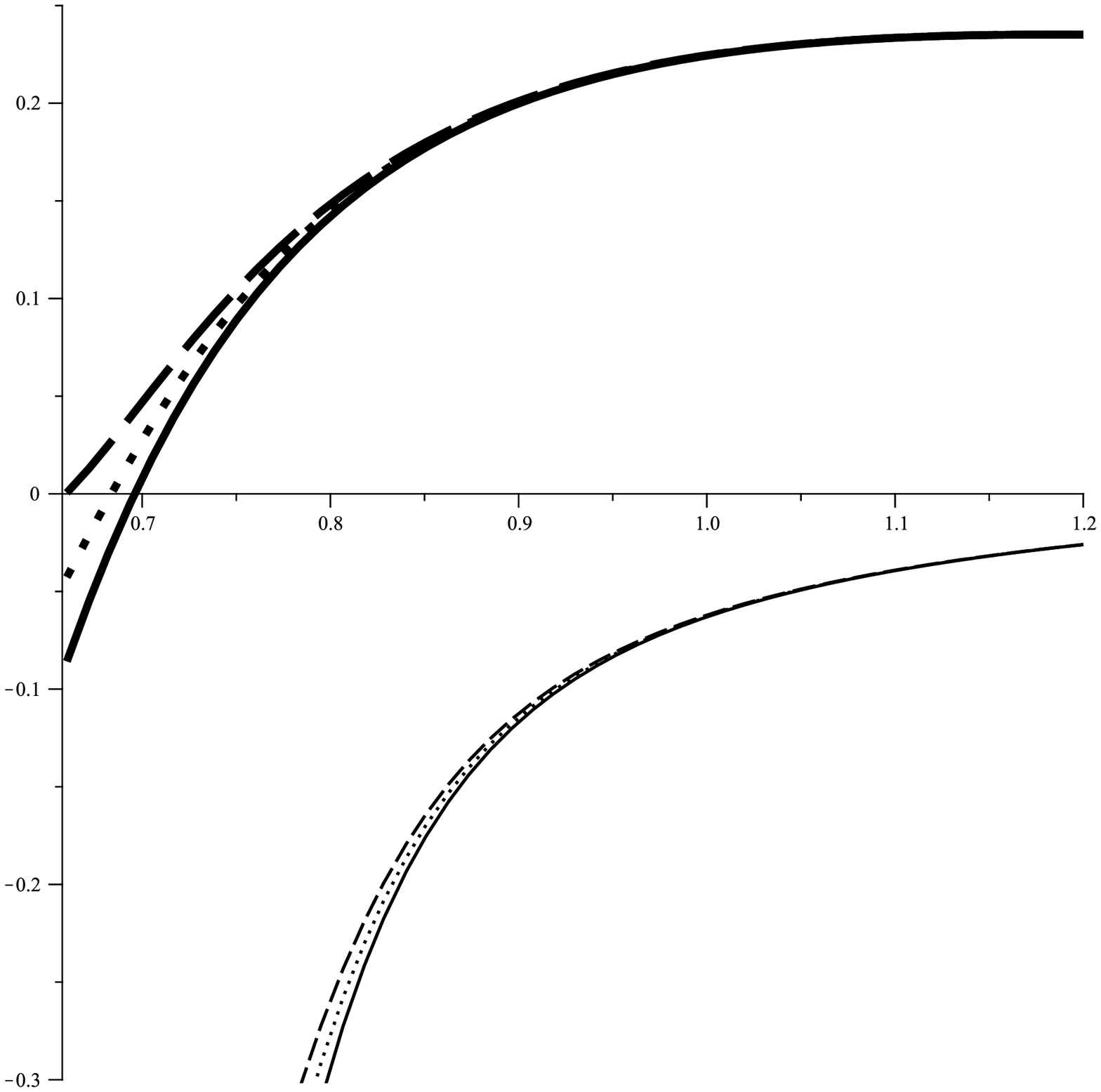}%
\end{array}
$%
\caption{\emph{Asymptotically flat solutions:} $\left( \frac{\partial ^{2}M}{%
\partial S^{2}}\right) _{Q}$ (left figure) and $\left\vert \mathbf{H}%
_{S,Q}^{M}\right\vert$ (right figure) versus $r_{+}$ for $n=5$, $q=1$, $%
\protect\alpha=0.1$ and $\protect\beta=0.001$ (solid line), $\protect\beta%
=0.003$ (dotted line) and $\protect\beta=0.005$ (dashed line).
\textbf{"bold lines represent the temperature"}}
\label{Stabilitybeta1}
\end{figure}

\begin{figure}[tbp]
$%
\begin{array}{cc}
\epsfxsize=7.5cm \epsffile{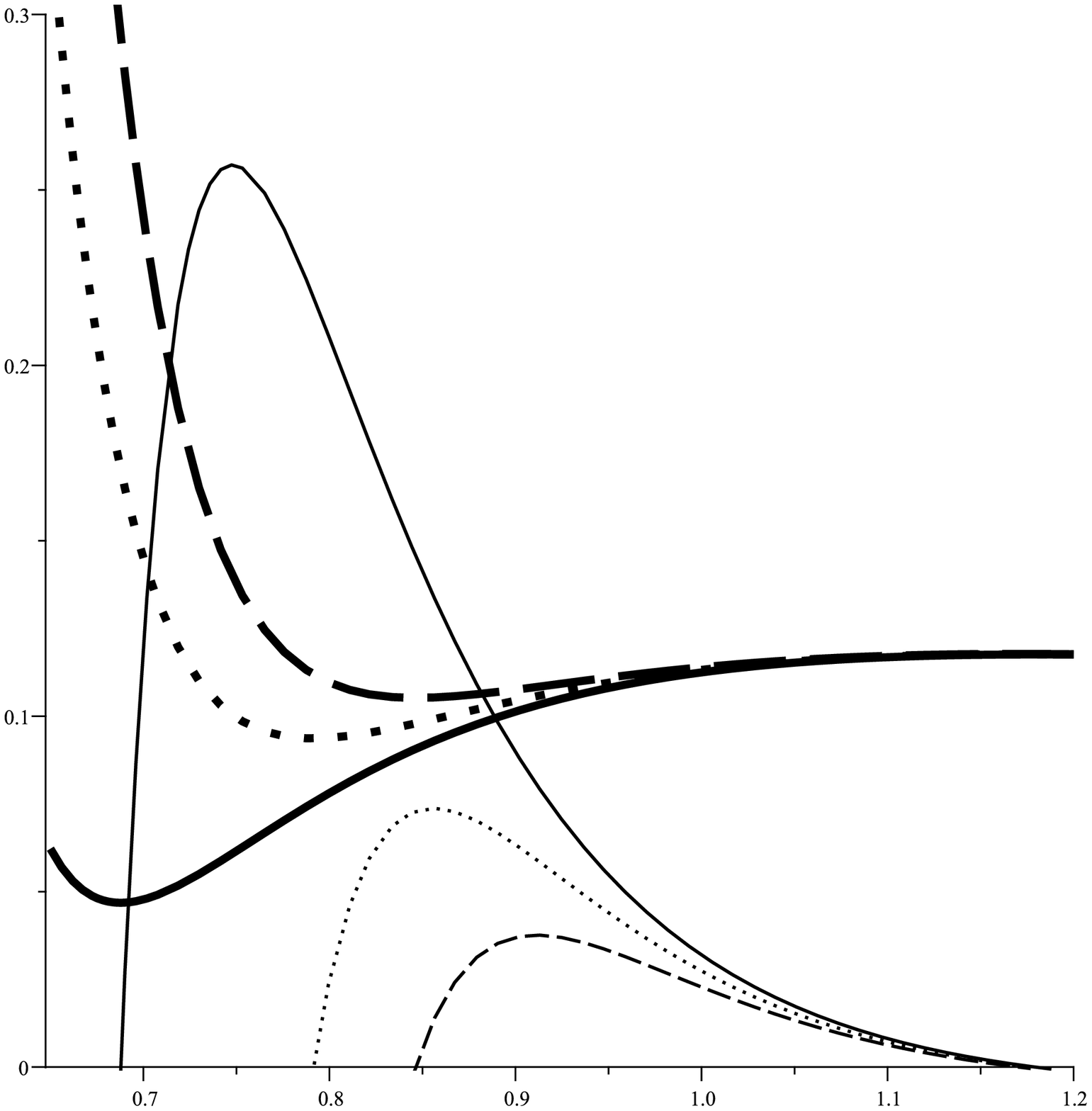} & \epsfxsize=7.5cm %
\epsffile{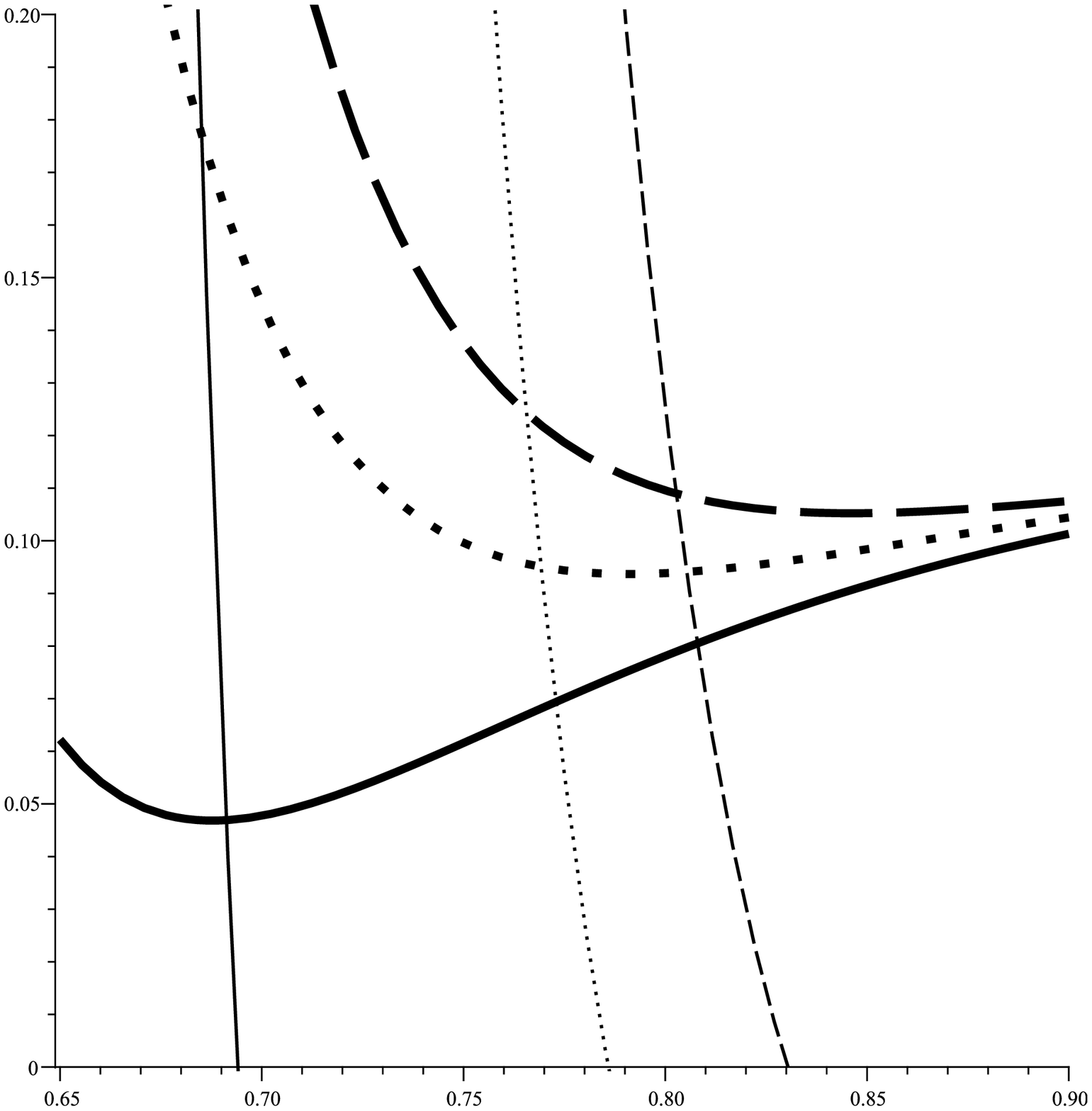}%
\end{array}
$%
\caption{\emph{Asymptotically flat solutions:} $\left( \frac{\partial ^{2}M}{%
\partial S^{2}}\right) _{Q}$ (left figure) and $\left\vert \mathbf{H}%
_{S,Q}^{M}\right\vert$ (right figure) versus $r_{+}$ for $n=5$, $q=1$, $%
\protect\alpha=0.1$ and $\protect\beta=0.01$ (solid line), $\protect\beta%
=0.03$ (dotted line) and $\protect\beta=0.05$ (dashed line).
\textbf{"bold lines represent the temperature"}}
\label{Stabilitybeta2}
\end{figure}

\begin{figure}[tbp]
$%
\begin{array}{cc}
\epsfxsize=7.5cm \epsffile{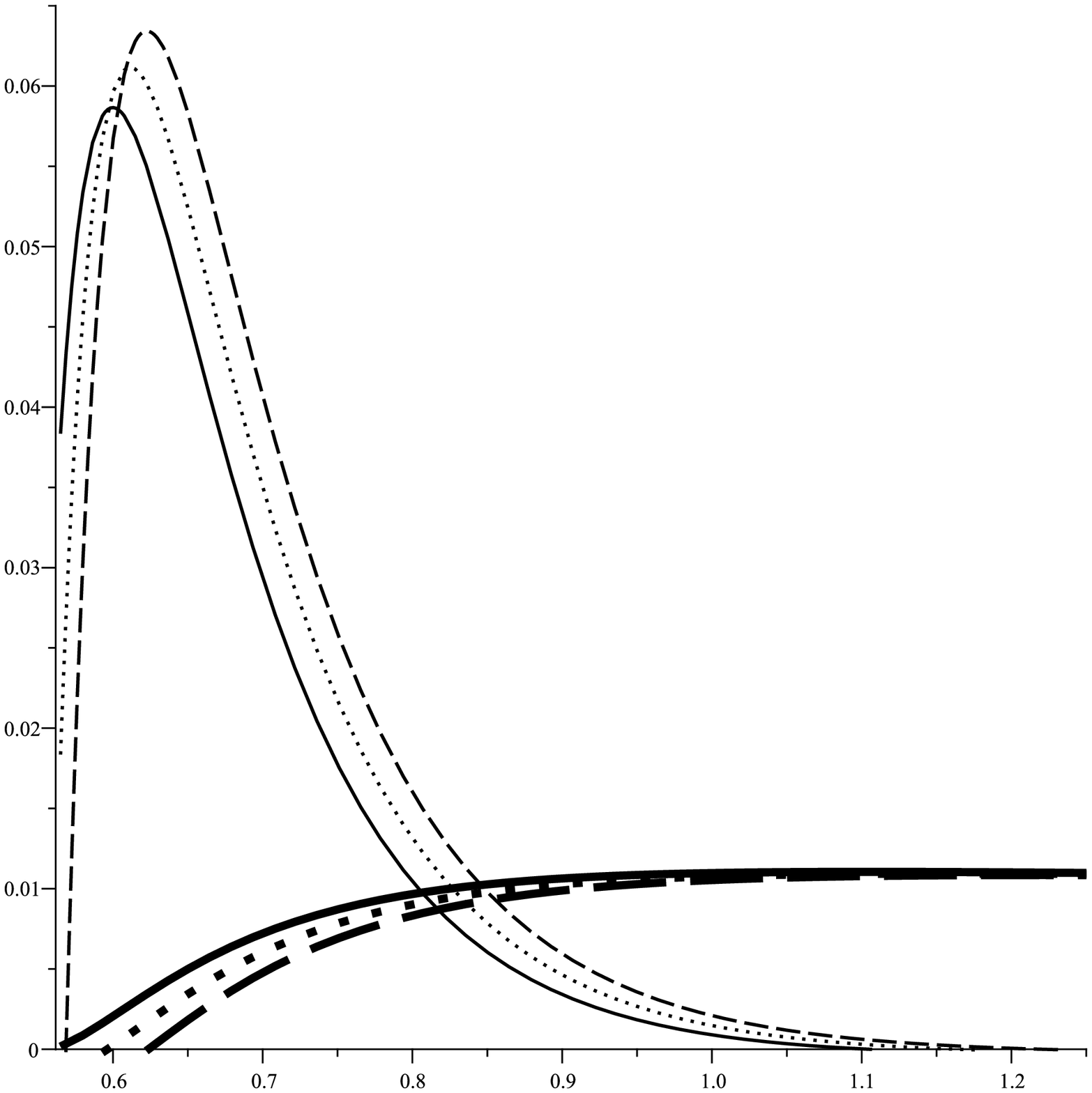} & \epsfxsize=7.5cm %
\epsffile{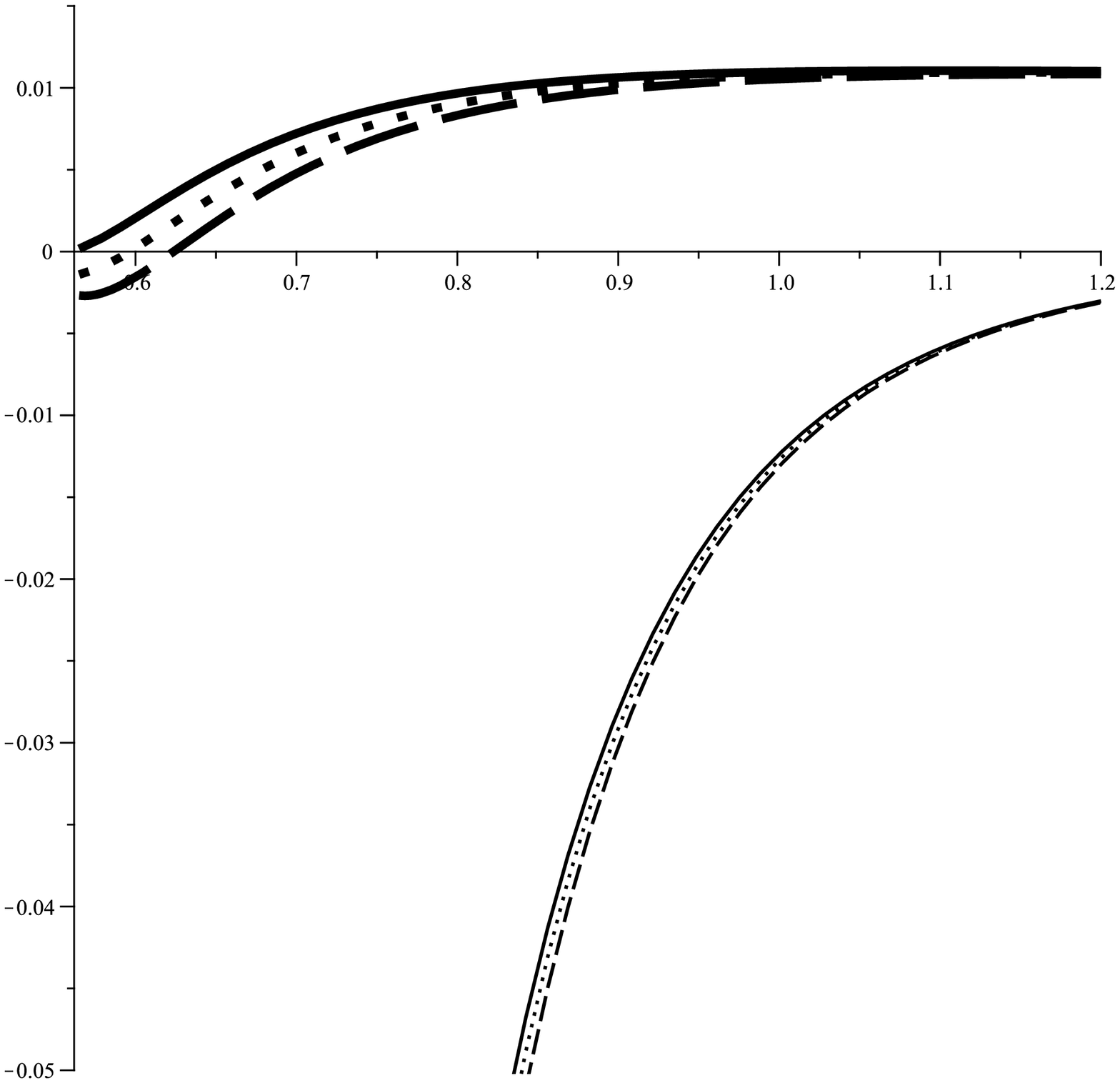}%
\end{array}
$%
\caption{\emph{Asymptotically flat solutions:} $\left( \frac{\partial ^{2}M}{%
\partial S^{2}}\right) _{Q}$ (left figure) and $\left\vert \mathbf{H}%
_{S,Q}^{M}\right\vert$ (right figure) versus $r_{+}$ for $n=5$, $\protect%
\beta=0.001$, $\protect\alpha=0.7$ and $q=1.2$ (solid line),
$q=1.3$ (dotted line) and $q=1.4$ (dashed line). \textbf{"bold
lines represent the temperature"}} \label{Stabilityq1}
\end{figure}

\begin{figure}[tbp]
$%
\begin{array}{cc}
\epsfxsize=7.5cm \epsffile{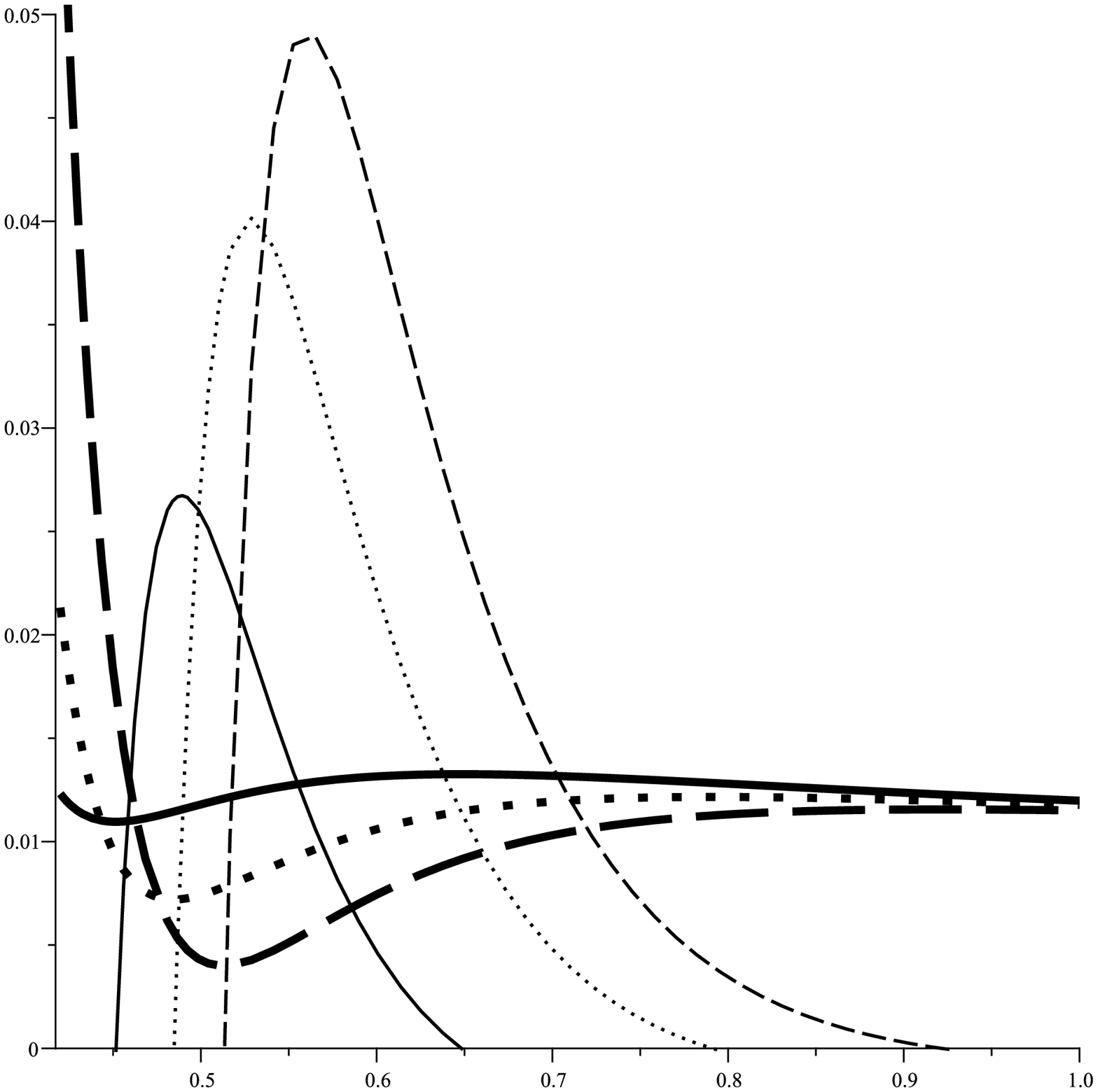} & \epsfxsize=7.5cm %
\epsffile{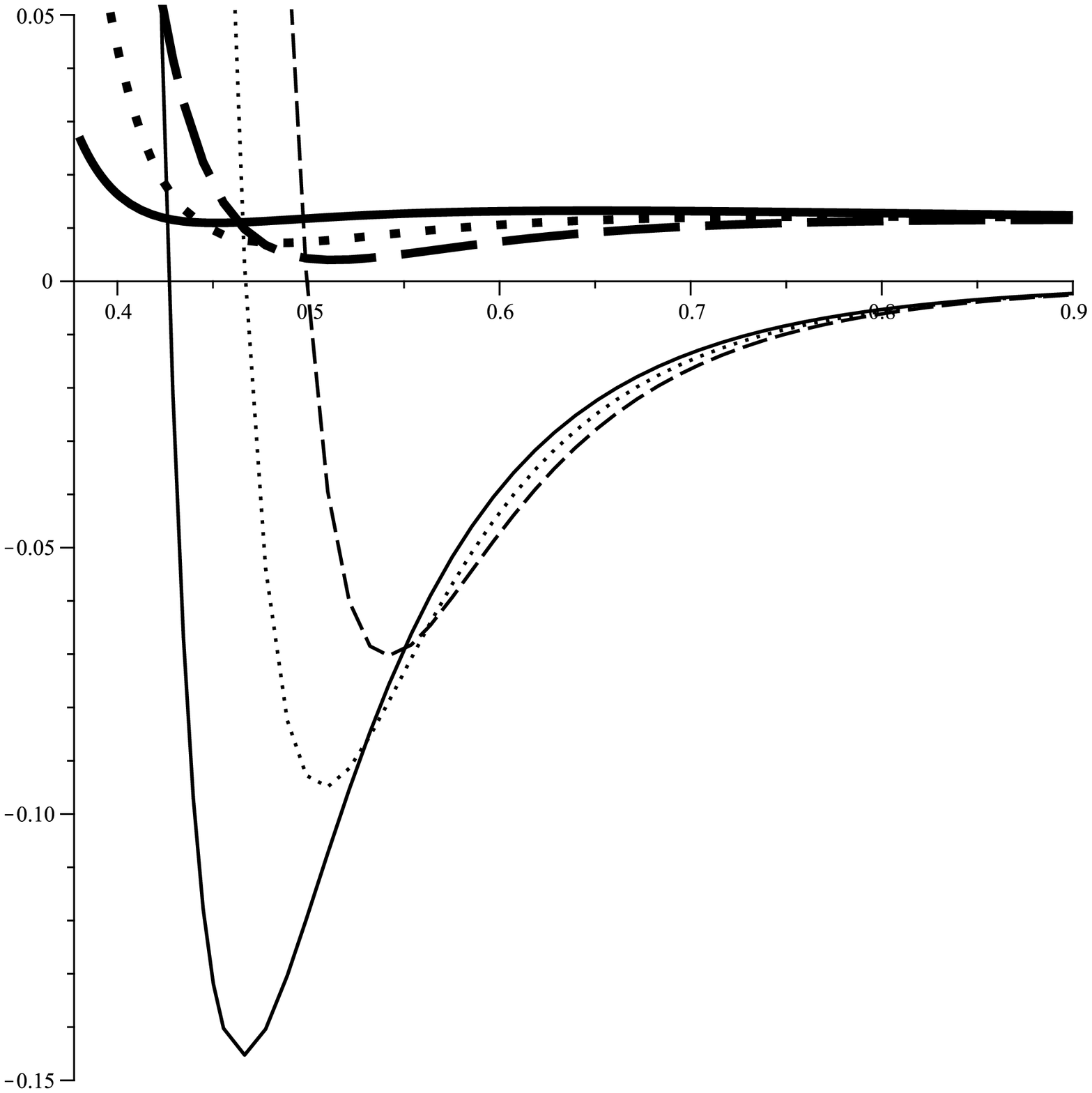}%
\end{array}
$%
\caption{\emph{Asymptotically flat solutions:} $\left( \frac{\partial ^{2}M}{%
\partial S^{2}}\right) _{Q}$ (left figure) and $\left\vert \mathbf{H}%
_{S,Q}^{M}\right\vert$ (right figure) versus $r_{+}$ for $n=5$, $\protect%
\beta=0.001$, $\protect\alpha=0.7$ and $q=0.5$ (solid line),
$q=0.7$ (dotted line) and $q=0.9$ (dashed line). \textbf{"bold
lines represent the temperature"}} \label{Stabilityq2}
\end{figure}

In order to investigate the effects of GB gravity and NLED on the
local stability for arbitrary $r_{+}$, we plot Figs.
\ref{Stabilityalpha1} -- \ref{Stabilityq2}. Here, we consider left
figures which are appropriate for the canonical ensemble analysis.

In Figs. \ref{Stabilityalpha1} and \ref{Stabilityalpha2} we investigate the
effects of considering GB gravity. One can find that when the GB parameter
is less than a critical value ($\alpha ^{\prime }<\alpha _{c}^{\prime }$), $%
T_{+}$ has a real positive root at $r_{0}$. In this case, $\left( \frac{%
\partial ^{2}M}{\partial S^{2}}\right) _{Q}>0$ for $r_{0}<r_{+}<r_{+\max}$.
In other words, small black holes are not physical and large black
holes are not stable. Moreover, we find that $r_{0}$($r_{+\max}$)
decrease (increase) when we increase $\alpha^{\prime}$. In Fig.
\ref{Stabilityalpha2}, we consider $\alpha^{\prime}>\alpha
_{c}^{\prime }$, in which $T_{+}$ is a positive definite function
of $r_{+}$. In this case one can obtain stable solutions for
$r_{+\min}<r_{+}<r_{+\max}$, in which confirms that small and
large black holes are unstable. We should note that increasing $%
\alpha^{\prime}$ leads to increasing (decreasing) the value of $r_{+\min}$ ($%
r_{+\max}$) and, therefore, decreases the range of stability.

Besides, we plot Figs. \ref{Stabilitybeta1} and
\ref{Stabilitybeta2} to analyze the effects of NLED. We find that
for the fixed values of $q$, $n$ and $\alpha^{\prime}$, there is a
critical value for the nonlinearity parameter, $\beta_{c}$, in
which for $\beta <\beta_{c}$, the temperature is positive for
$r_{+}>r_{0}$. This case is the same as that in Fig.
\ref{Stabilityalpha1} with $\alpha^{\prime} < \alpha_{c}^{\prime}$
and we may obtain asymptotically flat stable black holes for $
r_{0}<r_{+}<r_{+\max}$. In Fig. \ref{Stabilitybeta2} we set
$\beta>\beta_{c}$ to investigate the stability condition for
positive definite temperature. One finds there are lower and upper
limits for the horizon radius of stable black holes. It means that
for $\beta > \beta_{c}$, asymptotically flat black holes are
stable when $r_{+\min} < r_{+} < r_{+\max}$.

We study the effects of electric charge in Figs. \ref{Stabilityq1} and \ref%
{Stabilityq2}. These figures show that for fixed values of
$\alpha^{\prime}$ and $\beta$, there is a critical value for the
electric charge, $q_{c}$, in which the temperature is positive
definite for $q<q_{c}$. When $q>q_{c}$, there is a real positive
root ($r_{0}$) for the temperature which is an increasing function
of $q$. We should note that the general behaviors of Figs.
\ref{Stabilityq1} and \ref{Stabilityq2} are, respectively, the
same as those in Figs. \ref{Stabilityalpha1} and
\ref{Stabilityalpha2}. In other words, asymptotically flat black
holes are stable for $r_{+\min} < r_{+} < r_{+\max}$ and one
should replace $r_{+\min}$ with $r_{0}$ for $q>q_{c}$. It is
worthwhile to mention that both $r_{+\min}$ and $r_{+\max}$ are
increasing functions of $q$.

Finally, considering Figs. \ref{Stabilityalpha1} -- \ref{Stabilityq2}, we
should note that, in canonical ensemble, asymptotically flat black holes are
stable for $r_{+\min}<r_{+}<r_{+\max}$, in which one must replace $%
r_{+\min}$ with $r_{0}$ when $T_{+}$ has a real positive root
($q>q_{c}$ or $\beta<\beta_{c}$ or $\alpha^{\prime} <
\alpha_{c}^{\prime }$).

%%%%%%%%%%%%%%%%%%%%%%%%%%%%%%%%%%%%%%%%%%%%%%%%%%%%%%%%%%
%%%%%%%%%%%%%%%%%%%%%%%%%%%%%%%%%%%%%%%%%%%%%%%%%%%%%%%%%%

%%%%%%%%%%%%%%%%%%%%%%%%%%%%%%%%%%%%%%%%%%%%%%%%%%%%%%%%%%
%%%%%%%%%%%%%%%%%%%%%%%%%%%%%%%%%%%%%%%%%%%%%%%%%%%%%%%%%%

%%%%%%%%%%%%%%%%%%%%%%%%%%%%%%%%%%%%%%%%%%%%%%%%%%%%%%%%%%
%%%%%%%%%%%%%%%%%%%%%%%%%%%%%%%%%%%%%%%%%%%%%%%%%%%%%%%%%%
%%%%%%%%%%%%%%%%%%%%%%%%%%%%%%%%%%%%%%%%%%%%%%%%%%%%%%%%%%
%%%%%%%%%%%%%%%%%%%%%%%%%%%%%%%%%%%%%%%%%%%%%%%%%%%%%%%%%%

%%%%%%%%%%%%%%%%%%%%%%%%%%%%%%%%%%%%%%%%%%%%%%%%%%%%%%%%%%
%%%%%%%%%%%%%%%%%%%%%%%%%%%%%%%%%%%%%%%%%%%%%%%%%%%%%%%%%%
%%%%%%%%%%%%%%%%%%%%%%%%%%%%%%%%%%%%%%%%%%%%%%%%%%%%%%%%%%
%%%%%%%%%%%%%%%%%%%%%%%%%%%%%%%%%%%%%%%%%%%%%%%%%%%%%%%%%%

Next, in the grand-canonical ensemble, we calculate the
determinant of the Hessian matrix of the asymptotically flat black
holes. After some algebraic manipulation, one obtains
\begin{eqnarray}
\left\vert \mathbf{H}_{S,Q}^{M}\right\vert  &=&4\left\{
2q^{2}(3n-4)r_{+}^{4n-4}\left[ r_{+}^{2}-2\alpha ^{\prime }\right] +\right.
\nonumber \\
&&4q^{4}r_{+}^{2n-2}\left[ 2\alpha ^{\prime }(11n-24)-(7n-8)r_{+}^{2}\right]
\beta   \nonumber \\
&&\left. +12q^{2}(n-1)(n-2)^{2}r_{+}^{4n-8}\left[ \frac{(n-8)\alpha ^{\prime
}r_{+}^{2}}{(n-2)}\right. \right.   \nonumber \\
&&\left. \left. +\frac{2(n-4)\alpha ^{\prime 2}}{(n-2)}+r_{+}^{4}\right]
\beta \right.   \nonumber \\
&&\left. -(n-1)(n-2)(3n-4)r_{+}^{6n-10}\left[ \frac{(n-8)\alpha ^{\prime
}r_{+}^{2}}{(n-2)}\right. \right.   \nonumber \\
&&\left. \left. +\frac{2(n-4)\alpha ^{\prime 2}}{(n-2)}+r_{+}^{4}\right]
\right\} /\left[ r_{+}^{8n-14}\right.   \nonumber \\
&&\left. \left( r_{+}^{2}+2\alpha ^{\prime }\right)
^{3}(n-1)^{2}(3n-4)(n-2) \right] +O\left( \beta ^{2}\right) .
\label{Hessiank1}
\end{eqnarray}%
In addition, it is worthwhile to mention that for small and large
values of $r_{+}$, we obtain
\begin{eqnarray}
\left\vert \mathbf{H}_{S,Q}^{M}\right\vert _{\text{Small }r_{+}} &=&\left\{
\begin{array}{cc}
\frac{4(11n-24)\beta q^{4}}{\pi (n-1)^{2}(3n-4)\alpha ^{\prime
2}r_{+}^{6n-12}}>0, & \alpha ^{\prime }\neq 0,\beta \neq 0 \\
-\frac{16(7n-8)\beta q^{4}}{r_{+}^{6n-8}(n-1)^{2}(n-2)(3n-4)}<0, &
\alpha
^{\prime }=0,\beta \neq 0 \\
-\frac{2q^{2}}{(n-1)^{2}(n-2)\alpha ^{\prime 2}r_{+}^{4n-10}}<0, & \alpha
^{\prime }\neq 0,\beta =0 \\
\frac{8q^{2}}{(n-1)^{2}(n-2)r_{+}^{4n-6}}>0, & \alpha ^{\prime }=0,\beta =0%
\end{array}%
\right. ,  \label{lower2} \\
\left\vert \mathbf{H}_{S,Q}^{M}\right\vert _{\text{Large }r_{+}} &=&-\frac{4%
}{\pi (n-1)r_{+}^{n}}<0,  \label{upper2}
\end{eqnarray}%
which indicate that, for nonvanishing $\alpha ^{\prime }$ and
$\beta$, asymptotically flat black holes with a large horizon
radius are not stable. In other words, there is an upper limit,
$r_{+\max}$, for the asymptotically flat stable black holes
($r_{+}<r_{+\max}$). In the absence of the GB gravity case
($\alpha ^{\prime }=0,\beta \neq 0$), there is a lower limit for
stable solutions. It means that for $\alpha ^{\prime }=0$ the
results of the stability conditions for the canonical and grand
canonical ensembles are identical and both ensembles show that
small and large black holes are unstable. We should note that in
the presence of GB gravity thermal stability is ensemble
dependent. These results are shown in Figs. \ref{Stabilityalpha1}
-- \ref{Stabilityq2}.

Now, we need to discuss ensemble dependency \cite{Comer}. As we
know in the canonical ensemble the internal energy is allowed to
fluctuate with fixed electric charge and, therefore, the black
hole and the heat reservoir remain in thermal equilibrium with a
certain temperature. While in the grand canonical ensemble the
black hole is in both thermal and electrical equilibrium, with its
reservoir held at a constant temperature and a constant potential.
In other words, different boundary conditions lead to different
ensembles.

In the usual discussions of the stability criterion of black
holes, one expects ensemble independence of the system. Indeed,
different ensembles which imply different boundary conditions,
should lead to similar stability conditions in the usual
thermodynamical systems.

Ensemble dependency of a system may come from two subjects. One of
them is real ensemble dependency, which may occasionally occur in
some thermodynamical systems. The existence of ensemble dependency
was seen in the usual thermodynamical systems \cite{Usual}.
Another one, which is not real, can be removed by improving our
thermodynamic view point. In other words, our lack of knowledge
may lead to ensemble dependency and we should improve our
thermodynamical understanding to obtain ensemble independency. In
our black hole model (with a spherical horizon), we find that the
nonlinearity of electrodynamics results in the same changes for
the behavior of both canonical and grand canonical ensembles. One
can see the ensemble dependency comes from the contribution of
Gauss-Bonnet gravity. This means that in the absence of the GB
parameter, both ensembles have the same stability conditions. This
shows that the GB parameter makes more of a contribution to
thermodynamical behavior of the black hole systems. In other
words, one may consider the GB parameter not only as a fixed
parameter, but also as a thermodynamical parameter \cite{GBalpha}.
By doing so, the Hessian matrix for this system will be modified
and, therefore, we may expect to see that this modification solves
the ensemble dependency of the black hole system.

Another way to solve ensemble dependency is through considering
the fact that thermodynamical systems described by a
thermodynamical potential must be invariant under the Legendre
transformation. To do so, one can use the method that was
introduced in \cite{Quevedo1} and can use geometrothermodynamics
to solve ensemble dependency of the
system\cite{Quevedo1,Lifshitz}.

\section{Asymptotically adS Spinning Black Branes ($k=0$ \& $\Lambda \neq 0$)}

In this section, we will investigate rotating adS space-time. In
order to investigate the asymptotic behavior of the solutions for
$k=0$ and $\Lambda \neq 0$, one can use the series expansion of
the metric function for a large value of $r$. We obtain
\begin{equation}
f(r)=-\frac{2\Lambda _{\mathrm{eff}}}{n(n-1)}r^{2}-\frac{m}{\left( 1+\frac{%
4\alpha ^{\prime }\Lambda _{\mathrm{eff}}}{n(n-1)}\right) r^{n-2}}+\frac{%
2q^{2}}{(n-1)(n-2)\left( 1+\frac{4\alpha ^{\prime }\Lambda _{\mathrm{eff}}}{%
n(n-1)}\right) r^{2n-4}}+O\left( \frac{1}{r^{2n-2}}\right) ,
\label{f(r)k0bast}
\end{equation}%
where%
\begin{equation}
\Lambda _{\mathrm{eff}}=\frac{n(n-1)\left( \sqrt{1+\frac{8\alpha ^{\prime
}\Lambda }{n(n-1)}}-1\right) }{4\alpha ^{\prime }}.  \label{Lambdaeff}
\end{equation}%
Eq. (\ref{f(r)k0bast}) shows that for $k=0$ the solutions are
asymptotically adS with an effective cosmological constant
$\Lambda _{\mathrm{eff}}$. When constructing a rotating space-time
one can apply the following transformation in the static
space-time with $k=0$
\begin{eqnarray}
t &\longrightarrow &\Xi t-a_{i}\phi _{i},  \nonumber \\
\phi &\longrightarrow &\Xi \phi _{i}-\frac{a_{i}}{l}t.  \label{boost}
\end{eqnarray}

Using this transformation, the metric of $(n+1)$-dimensional
asymptotically adS rotating space-time with $p$ rotation
parameters is
\begin{eqnarray}
ds^{2} &=&-f(r)\left( \Xi dt-\sum\limits_{i=1}^{p}a_{i}d\phi _{i}\right)
^{2}+\frac{dr^{2}}{f(r)}+\frac{r^{2}}{l^{4}}\sum\limits_{i=1}^{p}\left(
a_{i}dt-\Xi l^{2}d\phi _{i}\right) ^{2}  \nonumber \\
&&-\frac{r^{2}}{l^{2}}\sum\limits_{i<j}^{p}\left( a_{i}d\phi _{i}-a_{j}d\phi
_{i}\right) ^{2}+r^{2}dX^{2},  \label{Metric2}
\end{eqnarray}%
where $\Xi =\sqrt{1+\sum_{i=1}^{p}\frac{a_{i}}{l^{2}}}$ and
$dX^{2}$ is the Euclidean metric on the $(n-1-p)$-dimensional
submanifold. The forth term in Eq. (\ref{Metric2}) comes from the
fact that for rotating space-time with more than one rotating
parameter, we should consider cross terms associated with rotating
coordinates. The rotation group in $(n+1)$ dimensions is $SO(n)$
and, therefore, $p\leq \lbrack n/2]$. Considering the
aforementioned transformation, for the gauge potential, we should
write
\begin{equation}
A_{\mu }=h(r)\left( \Xi \delta _{\mu }^{0}-a_{i}\delta _{\mu }^{i}\right)
\text{ \ \ \ (no sum on }i\text{).}  \label{Amu}
\end{equation}%
Calculations show that metric function (\ref{f(r)}) with $k=0$ and $%
h(r)=\int F_{tr}dr$ (using the $F_{tr}$ calculated in Eq.
(\ref{Ftr})) satisfies all of the field equations for the
aforementioned rotating adS spacetime. Straightforward
calculations confirm that there is a curvature singularity located
at $r=0$ which is covered with the horizon(s).

Using the analytic continuation of the metric by setting $t\rightarrow i\tau
$ and $a_{i}\rightarrow ia_{i}$, and regularity at the event horizon, $r_{+}$%
, helps us to obtain the Hawking temperature and the angular
velocities of the black branes:
\begin{equation}
T_{+}=\frac{f^{\prime }\left( r_{+}\right) }{4\pi \Xi }=-\frac{%
r_{+}^{4n-4}\Lambda +r_{+}^{2n-2}q^{2}-2q^{4}\beta }{2\pi \Xi (n-1)r_{+}^{4n-5}}%
+O(\beta ^{2}),  \label{Tk0}
\end{equation}%
\begin{equation}
\Omega _{i}=\frac{a_{i}}{\Xi l^{2}}.  \label{Omega}
\end{equation}%
Eq. (\ref{Tk0}) shows that, unlike black hole solutions with a
spherical horizon, the temperature of the aforementioned black
brane with a flat horizon does not depend on GB gravity.

Next, we calculate the electric charge and potential of the solutions. The
electric charge per unit volume $V_{n-1}$ can be found by calculating the
flux of the electric field at infinity, yielding
\begin{equation}
Q=\frac{q\Xi }{4\pi }.  \label{Qk0}
\end{equation}

On the other hand, because of the applied transformation and
changes in metric we now have Killing vectors in the form of $\chi
=\partial _{t}+\sum_{i=1}^{p}\Omega _{i}\partial _{\phi _{i}}$
which is the null generator of the horizon. The electric potential
is obtained as
\begin{equation}
\Phi =\left. A_{\mu }\chi ^{\mu }\right\vert _{r\longrightarrow \infty
}-\left. A_{\mu }\chi ^{\mu }\right\vert _{r\longrightarrow r_{+}}=\frac{q}{%
\Xi (n-2)r_{+}^{n-2}}\left( 1-\frac{4\left( n-2\right) q^{2}\beta
}{\left( 3n-4\right) r_{+}^{2n-2}}\right) +O(\beta ^{2}).
\label{Phik0}
\end{equation}

Now, we desire to calculate the entropy, the angular momentum and
the finite mass to check the first law of thermodynamics. In
general, the action and the conserved quantities of the space-time
diverge when evaluated on the solutions. In order to overcome this
problem and due to the fact that our space-time is asymptotically
adS, we can use the counterterm method to calculate the finite
action and the conserved quantities. One can show that for the
obtained solutions with flat boundary, $R_{abcd}(\gamma )=0$, the
finite action is
\begin{equation}
I_{\mathrm{finite}}=I_{G}+I_{b}+I_{ct},  \label{Itot}
\end{equation}%
where
\begin{eqnarray}
I_{G} &=&-\frac{1}{16\pi }\int_{\mathcal{M}}d^{n+1}x\sqrt{-g}L_{\mathrm{tot}}
\label{IG} \\
I_{b} &=&-\frac{1}{8\pi }\int_{\partial \mathcal{M}}d^{n}x\sqrt{-\gamma }%
\left\{ K+2\alpha \left( J-2\hat{G}_{ab}K^{ab}\right) \right\}  \label{Ib} \\
I_{ct} &=&-\frac{1}{8\pi }\int_{\partial \mathcal{M}}d^{n}x\sqrt{-\gamma }%
\left( \frac{n-1}{l_{\mathrm{eff}}}\right) ,  \label{Ict}
\end{eqnarray}%
where $\gamma$ and $K$ are, respectively, the trace of the induced metric, $%
\gamma _{ab}$ and the extrinsic curvature, $K^{ab}$, on the boundary $%
\partial \mathcal{M}$, $\hat{G}_{ab}$\ is the Einstein tensor calculated on
the boundary, $J$ is the trace of
\begin{equation}
J_{ab}=\frac{1}{3}%
(K_{cd}K^{cd}K_{ab}+2KK_{ac}K_{b}^{c}-2K_{ac}K^{cd}K_{db}-K^{2}K_{ab}),
\label{Jab}
\end{equation}%
and $l_{\mathrm{eff}}$\ is a scale length factor that depends on $l$ and $%
\alpha $, which must reduce to $l$ as $\alpha $ vanishes. It is
worthwhile to mention that, for a spacetime with zero curvature
boundary, $I_{ct}$ has exactly the same value as that of the
Einstein gravity in which $l$ is replaced by $l_{\mathrm{eff}}$.
Using the Brown--York method of a quasilocal definition with Eq.
(\ref{Itot})-(\ref{Ict}), one can introduce a divergence-free
stress-energy tensor as follows
\begin{equation}
T^{ab}=\frac{1}{8\pi }\left[ (K^{ab}-K\gamma ^{ab})+2\alpha (3J^{ab}-J\gamma
^{ab})+\frac{n-1}{l_{\mathrm{eff}}}\gamma ^{ab}\right] .  \label{Tab}
\end{equation}

Then the quasilocal conserved quantities associated with the stress tensors
of Eq. (\ref{Tab}) can be written as
\begin{equation}
Q(\xi )=\int\limits_{B}d^{n-1}\varphi \sqrt{\gamma }T_{ab}n^{a}\xi ^{b},
\label{Conserved}
\end{equation}%
where $n^{a}$ is the timelike unit normal vector to the boundary $B$. Taking
into account Eq. (\ref{Conserved}) with $\xi =\partial /\partial t$ as a
Killing vector, one can calculate the mass per unit volume $V_{n-1}$ as
\begin{equation}
M=\frac{1}{16\pi }m\left( n\Xi ^{2}-1\right) ,  \label{Massk0}
\end{equation}%
where
\begin{equation}
m=m(r=r_{+}).  \label{mh}
\end{equation}%
Eqs. (\ref{Massk0}) and (\ref{mh}) indicate that, unlike the
asymptotic flat solutions with a spherical boundary, the finite
mass does not depend on the GB parameter for the boundary flat
rotating solutions.

Considering another Killing vector $\zeta =\partial /\partial \phi
_{i}$ of rotating spacetime which is related to angular momentum,
one can calculate the angular momentum per unit volume,
\begin{equation}
J_{i}=\frac{1}{16\pi }nm\Xi a_{i}.  \label{J}
\end{equation}

We should note that for static solutions ($a_{i}=0$), the angular
momentum vanishes, which confirms that the $a_{i}$'s are the
rotational parameters of the space-time.

As we mentioned before, the Wald formula can be applied to
asymptotically flat space-time. Here, we encounter asymptotically
adS solutions, so we calculate the entropy through the use of the
Gibbs-Duhem relation,
\begin{equation}
S=\frac{1}{T}\left( M-Q\Phi -\sum\limits_{i=1}^{k}\Omega _{i}J_{i}\right)
-I_{\mathrm{finite}}.  \label{GD}
\end{equation}%
First we calculate the finite action $I_{\mathrm{finite}}$ for the rotating
metric. We find that the finite action per unit volume may be written as
\begin{equation}
I_{\mathrm{finite}}=\frac{1}{8 \pi n(n-1) T_{+}} \left(   -\Lambda r_{+}^{n}+\frac{n q^{2}}{%
(n-2)r_{+}^{n-2}}-\frac{2 n q^{4}\beta }{(3n-4)r_{+}^{3n-4}}%
+O(\beta ^{2}) \right). \label{Ifinite}
\end{equation}

Using the finite conserved and thermodynamic quantities with the
finite action, we obtain
\begin{equation}
S=\frac{\Xi }{4}r_{+}^{n-1},  \label{entropyk0}
\end{equation}
which confirms that the entropy obeys the area law for asymptotically adS
black branes with zero curvature horizon.

Now we want to check the first law of thermodynamics. To do so, we
obtain the mass as a function of the extensive quantities $S$, $J$
and $Q$. Using the expression for the electric charge, the mass,
the angular momentum and
the entropy given, respectively, in Eqs. (\ref{Qk0}), (\ref{Massk0}), (\ref%
{J}) and (\ref{entropyk0}) and the fact that $f(r=r_{+})=0$, one can obtain
a Smarr-type formula as
\begin{equation}
M\left( S,J\mathbf{,}Q\right) \mathbf{=}\frac{\left( nZ-1\right) J}{nl\sqrt{%
Z\left( Z-1\right) }},  \label{Msmark0}
\end{equation}%
where $J=|\mathbf{J}|=\sqrt{\sum_{i=1}^{p}J_{i}^{2}}$ and $Z=\Xi ^{2}$ is
the positive real root of the following equation
\begin{eqnarray}
&&2^{\frac{2}{n-1}}nl\sqrt{Z\left( Z-1\right) }\left( \pi ^{4}Q^{4}\beta -%
\frac{\left( 3n-4\right) S^{2}\pi ^{2}Q^{2}}{2(n-2)}+\frac{\left(
3n-4\right) \Lambda S^{4}}{2n}\right)+  \nonumber \\
&&\left( \frac{S}{\sqrt{Z}}\right) ^{\frac{n-2}{n-1}}S^{2}Z\pi \left(
n-1\right) \left( 3n-4\right) J=0.  \label{Zeq}
\end{eqnarray}

Taking into account $S$, $J$ and $Q$ as the extensive parameters
of $M$, we can define the intensive parameters conjugate to them
as
\begin{equation}
T=\left( \frac{\partial M}{\partial S}\right) _{J,Q}\ \ ,\ \ \Omega
_{i}=\left( \frac{\partial M}{\partial J_{i}}\right) _{S,Q}\ \ ,\ \ \Phi
=\left( \frac{\partial M}{\partial Q}\right) _{S,J}  \label{TOmegaPhik0}
\end{equation}

It is a matter of calculation to show that the intensive
quantities calculated by Eq. (\ref{TOmegaPhik0}) coincide with
Eqs. (\ref{Tk0}), (\ref{Omega}) and (\ref{Phik0}). Therefore, the
first law of thermodynamics is satisfied
\begin{equation}
dM=TdS+\sum\limits_{i}^{p}\Omega _{i}dJ_{i}+\Phi dQ.  \label{Firstk0}
\end{equation}

\subsection{Stability of the solutions}

Now, we are in a position to calculate the heat capacity and the
Hessian matrix to check the local stability of these solutions in
context of canonical and grand canonical ensembles. For canonical
ensembles, where electric charge
and angular momenta are fixed parameters, the positivity of $\left( \frac{%
\partial ^{2}M}{\partial S^{2}}\right) _{J,Q}$ is sufficient to ensure the
local stability. Therefore, we obtain
\begin{equation}
\left( \frac{\partial ^{2}M}{\partial S^{2}}\right) _{J,Q}=\frac{%
2r_{+}^{-2}A_{1}}{\pi \Xi ^{2}\left( n-1\right) \psi \sigma }-\frac{%
4(n-2)^{2}q^{4}r_{+}^{-4}A_{2}\beta }{\pi \Xi ^{2}\left(
n-1\right) \left( 3n-4\right) \sigma \psi ^{2}}+O\left( \beta
^{2}\right) ,  \label{dMdSSk0}
\end{equation}

\begin{equation}
A_{1}=\frac{n\left( 3\sigma -n\right)
q^{4}}{r_{+}^{3n-6}}+\frac{2\left( 3\sigma -n^{2}\right) \Lambda
q^{2}}{r_{+}^{n-4}}+\frac{\left[ (n+2)\sigma -n^{2}\right] \Lambda
^{2}}{r_{+}^{2-n}} , \label{A1}
\end{equation}

\begin{eqnarray}
A_{2} &=&\frac{\left[ (5n^{2}-42n+40)\Xi ^{2}+(7n^{2}+11n-20)\right] \Lambda
^{2}}{r_{+}^{n-6}}-  \nonumber \\
&&\frac{2n\left[ (13n^{2}-50n+40)\Xi ^{2}-(n^{2}-19n+20)\right] \Lambda q^{2}%
}{(n-2)r_{+}^{3n-8}}+  \nonumber \\
&&\frac{n^{2}\left[ (17n-20)\sigma -n(5n-6)\right] q^{4}}{%
(n-2)^{2}r_{+}^{5n-10}},  \label{A2}
\end{eqnarray}

\begin{eqnarray}
\psi  &=&\left[ nq^{2}-\Lambda (n-2)r_{+}^{2n-2}\right] , \\
\sigma  &=&\left[ (n-2)\Xi ^{2}+1\right].
\end{eqnarray}

As we mentioned before, in the grand canonical ensemble, the positivity of
the determinant of the Hessian matrix of $M(S,Q,J)$ with respect to its
extensive variables $X_{i}$, $\mathbf{H}_{X_{i}X_{j}}^{M}$ $=\left( \frac{%
\partial ^{2}M}{\partial X_{i}\partial X_{j}}\right) $, is sufficient to
ensure the local stability. It is a matter of calculation to show that the
determinant of $\mathbf{H}_{S,Q,J}^{M}$ is
\begin{eqnarray}
\left\vert \mathbf{H}_{A,Q,J}^{M}\right\vert  &=&\varkappa (q^{2}-\Lambda
r_{+}^{2n-2})-  \nonumber \\
&&\frac{4\varkappa q^{2}\left[ 3(n-2)^{2}\Lambda
^{2}r_{+}^{4n-4}+3n(n-1)q^{4}-2(n-2)(3n-2)\Lambda
q^{2}r_{+}^{2n-2}\right] \beta }{(3n-4)\left[ nq^{2}-\left(
n-2\right) \Lambda r_{+}^{2n-2}\right] r_{+}^{2n-2}}+O(\beta
^{2}),  \label{Hessiank0}
\end{eqnarray}%
where%
\begin{equation}
\varkappa =\frac{64\pi r_{+}^{4-3n}\left[ nq^{2}-\left( n-2\right) \Lambda
r_{+}^{2n-2}\right] ^{-1}}{\left[ \left( n-2\right) \Xi ^{2}+1\right] \Xi
^{6}l^{2}}.  \label{kap}
\end{equation}

Regarding Eq. (\ref{Tk0}) with the mentioned $\left( \frac{\partial ^{2}M}{%
\partial S^{2}}\right) _{J,Q}$ and also $\left\vert \mathbf{H}%
_{A,Q,J}^{M}\right\vert $, we find that, unlike the asymptotically
flat case with the spherical horizon, neither the heat capacity
nor the determinant of the Hessian matrix depend on the GB
parameter for asymptotically adS rotating solutions with zero
curvature horizon. Therefore, we expect to obtain the same results
in both canonical and grand canonical ensembles. Following the
method of the previous section and regardless of the values of
$n$, $q$, $\Lambda $, $\Xi $ and $\alpha $, one finds
\begin{eqnarray}
\left. \left( \frac{\partial ^{2}M}{\partial S^{2}}\right) _{J,Q}\right\vert
_{\text{Small }r_{+}} &=&\left\{
\begin{array}{cc}
-\frac{4\left[ (5n^{2}-23n+20)\left( \Xi ^{2}-1\right)
+(12n^{2}-31n+20)\Xi ^{2}\right] q^{4}\beta }{\pi \Xi ^{2}\left(
n-1\right) \left( 3n-4\right)
\sigma r_{+}^{5n-6}}<0, & \beta \neq 0 \\
\frac{2\left\{ 3\left[ (n-2)\Xi ^{2}+1\right] -n\right\} q^{2}}{\pi \Xi
^{2}\left( n-1\right) \left[ (n-2)\Xi ^{2}+1\right] r_{+}^{3n-4}}>0, &
\beta =0%
\end{array}%
\right. ,  \label{lower3} \\
\left. \left( \frac{\partial ^{2}M}{\partial S^{2}}\right) _{J,Q}\right\vert
_{\text{Large }r_{+}} &=&-\frac{2\left[ (n^{2}-4)\left( \Xi ^{2}-1\right)
+n-2\right] \Lambda }{\pi \Xi ^{2}\left( n-1\right) (n-2)\left[ (n-2)\Xi
^{2}+1\right] r_{+}^{n+2}}>0,  \label{upper3}
\end{eqnarray}%
and
\begin{eqnarray}
\left. H_{S,Q,J}^{M}\right\vert _{\text{Small }r_{+}} &=&\left\{
\begin{array}{cc}
-\frac{768\pi (n-1)q^{2}\beta }{n(3n-4)\left[ \left( n-2\right) \Xi ^{2}+1%
\right] \Xi ^{6}l^{2}r_{+}^{5n-6}}<0, & \beta \neq 0 \\
\frac{64\pi }{n\left[ \left( n-2\right) \Xi ^{2}+1\right] \Xi
^{6}l^{2}r_{+}^{3n-4}}>0, & \beta =0%
\end{array}%
\right. ,  \label{lower4} \\
\left. H_{S,Q,J}^{M}\right\vert _{\text{Large }r_{+}} &=&\frac{64\pi }{%
(n-2)l^{2}\Xi ^{6}\left[ \left( n-2\right) \Xi ^{2}+1\right] r_{+}^{3n-4}}>0.
\label{upper4}
\end{eqnarray}%
Both ensembles confirm that, in the presence of NLED ($\beta \neq
0$), although the black branes with small $r_{+}$ are unstable,
large black branes are stable. It is notable that the instability
of the small black branes is due to the presence of the NLED and,
in the absence of the nonlinearity effect, small black branes are
stable.

\section{Closing Remarks}
In this paper, we regarded both the gravity and electrodynamic
string corrections of Einstein-Maxwell gravity to obtain black
hole solutions with spherical, hyperbolic, and flat horizon
topology.

At the first step, we focused on asymptotically flat solutions. We
used the Wald formula, the Gauss law and the ADM approach to
calculate entropy, electric charge and finite mass, respectively.
We checked that the conserved and thermodynamic quantities
satisfied the first law of thermodynamics. Then we investigated
the thermodynamic stability of the solutions in both canonical and
grand canonical ensembles. Taking into account the canonical
ensemble, we found that for nonzero $\alpha^{\prime }$ and
$\beta$, asymptotically flat black holes with a large or small
horizon radius are unstable. This means that, in canonical
ensemble, asymptotically flat black holes are stable for $r_{+\min
}<r_{+}<r_{+\max}$, in which one must replace $r_{+\min}$ with
$r_{0}$ (the largest root of $T_{+}$) when $T_{+}$ has a real
positive root ($q>q_{c}$ or $\beta<\beta_{c}$ or $\alpha^{\prime}
< \alpha_{c}^{\prime }$). Moreover, we found that different values
of $\alpha$, $\beta$ and $q$ can change the values of $r_{+\min}$
and $r_{+\max}$.

Then, we investigated the stability conditions in the grand
canonical ensemble. We showed that there is an upper limit,
$r_{+\max}$, for the asymptotically flat stable black holes
($r_{+}<r_{+\max}$). We found that, although for $\alpha^{\prime
}=0$ the results of the stability conditions for canonical and
grand canonical ensembles are identical, these ensembles have
different consequences in the presence of GB gravity. In other
words, we noted that in the presence of GB gravity, thermal
stability is ensemble dependent.

In the next section, we considered the Ricci flat solutions with
an adS asymptote and produced a rotating spacetime by using an
improper local transformation. In addition, we calculated the
conserved and thermodynamic quantities for asymptotically adS
black branes which satisfy the first law of thermodynamics.
Considering the thermodynamic instability criterion in canonical
and grand canonical ensembles, we found that neither the heat
capacity nor the Hessian matrix depend on the GB parameter.
Therefore, both ensembles have identical stability conditions. We
found that, although the black branes with small $r_{+}$ are
unstable, large black branes are stable (unlike asymptotically
flat static large black holes which are unstable). It is notable
that the instability of small black branes (holes) is due to the
presence of NLED and, in the absence of the nonlinearity effect,
small black branes (holes) are stable.

To conclude, we found that for the Ricci-flat solutions, we
obtained the same conditions for stable black holes in both
canonical and grand canonical ensembles. In other words, in this
case, we took into account $S$, $Q$ and $J$ as the thermodynamical
extensive parameters, correctly, and obtained the same results for
both ensembles. It is easy to show that, for rotating Ricci-flat
solutions, one may obtain ensemble dependency if one, imprecisely,
considers $S$ and $Q$ as the set of extensive parameters (regards
$J$ as a dynamical parameter, not a thermodynamic one). As one can
confirm, unlike Ricci-flat solutions, the GB parameter in the
spherical horizon of GB black holes contributes to finite mass,
temperature and, consequently, heat capacity. In other words, the
GB parameter contributions lead to ensemble dependency in
spherical horizon black holes. This means that a GB generalization
of Einstein gravity not only affects gravitational properties, but
also thermodynamic aspects of the spherically symmetric black
holes.

Finally, we should note that the modifications of the
thermodynamic instability criterion in the presence of the GB
gravity depend on the choice of the ensemble. One may consider the
GB parameter as a thermodynamic variable to remove the ensemble
dependency. In addition, it is worthwhile to mention that we can
regard asymptotically adS black holes with spherical topology to
investigate $P-V$ criticality in the extended phase space of the
solutions by calculating the Gibbs free energy. These works are
under examination.

\begin{acknowledgements}
We thank the Shiraz University Research Council. This work has
been supported financially by the Research Institute for Astronomy
and Astrophysics of Maragha, Iran.
\end{acknowledgements}


\begin{thebibliography}{99}
\bibitem{Born} M. Born and L. Infeld, Proc. Roy. Soc. Lond. A 143, 410
(1934);

M. Born and L. Infeld, Proc. Roy. Soc. Lond. A 144, 425 (1934).

\bibitem{BIpapers} D. L. Wiltshire, Phys. Rev. D 38, 2445 (1988);

M. Cataldo and A. Garcia, Phys. Lett. B 456, 28 (1999);

S. Fernando and D. Krug, Gen. Relativ. Gravit. 35, 129 (2003);

T. Tamaki, J. Cosmol. Astropart. Phys. 05, 004 (2004);

M. Aiello, R. Ferraro, and G. Giribet, Phys. Rev. D 70, 104014
(2004);

T. K. Dey, Phys. Lett. B 595, 484 (2004);

R. G. Cai, D.W. Pang, and A. Wang, Phys. Rev. D 70, 124034
(2004);

M. H. Dehghani and H. R. Rastegar-Sedehi, Phys. Rev. D 74, 124018
(2006);

M.H. Dehghani, N. Alinejadi, and S. H. Hendi, Phys. Rev. D 77,
104025 (2008);

S. H. Hendi, J. Math. Phys. (N.Y.) 49, 082501 (2008).


\bibitem{PMIpapers} M. Hassaine and C. Martinez, Phys. Rev. D 75, 027502
(2007);

M. Hassaine and C. Martinez, Class. Quantum Gravit. 25, 195023
(2008);

M. Maeda, M. Hassaine and C. Martinez, Phys. Rev. D 79, 044012 (2009).

\bibitem{Soleng} H. H. Soleng, Phys. Rev. D 52, 6178 (1995).

\bibitem{HendiJHEP} S. H. Hendi, JHEP 03, 065 (2012).

\bibitem{HendiNLED} S. H. Hendi, Ann. Phys. (N.Y.) 333, 282 (2013);

S. H. Hendi and A. Sheykhi, Phys. Rev. D 88, 044044 (2013).

\bibitem{BItypeProperties} G. Boillat, J. Math. Phys. 11, 941 (1970);

G. Boillat, J. Math. Phys. 11, 1482 (1970);

G. W. Gibbons and D. A. Rasheed, Nucl. Phys. B 454, 185 (1995);

Z. Zhao, Q. Pan, S. Chen and J. Jing, Nucl. Phys. B 871, 98
(2013).

\bibitem{BItyprString} E. S. Fradkin and A. A. Tseytlin, Phys. Lett. B 163,
123 (1985);

J. Ambjorn, Y. M. Makeenko, J. Nishimura and R. J. Szabo, Phys. Lett. B 480,
399 (2000);

N. Seiberg and E. Witten, JHEP 09, 032 (1999);

R. Matsaev, M. Rahmanov and A. Tseytlin, Phys. Lett. B 193, 207
(1987);

E. Bergshoff, E. Sezgin, C. Pope and P. Townsend, Phys. Lett. B 188, 70
(1987);

Y. Kats, L. Motl and M. Padi, JHEP 12, 068 (2007);

R. G. Cai, Z. Y. Nie and Y. W. Sun, Phys. Rev. D 78, 126007 (2008);

D. Anninos and G. Pastras, JHEP 07, 030 (2009).

\bibitem{HendiMomennia} S. H. Hendi and M. Momennia, \emph{submitted for publication}.

\bibitem{HendiEPJC} S. H. Hendi, Eur. Phys. J. C 73, 2634 (2013).

\bibitem{HPphase} S. W. Hawking and D. N. Page, Commun. Math. Phys. 87, 577
(1983).

\bibitem{SchPhase} D. Birmingham, Class. Quantum Gravit. 16, 1197 (1999).

\bibitem{GBNLED} M. H. Dehghani and S. H. Hendi, Int. J. Mod. Phys. D 16, 1829
(2007);

O. Miskovic and R. Olea, Phys. Rev. D 83, 024011 (2011);

O. Miskovic and R. Olea, Phys. Rev. D 83, 064017 (2011);

S. H. Hendi and B. Eslam Panah, Phys. Lett. B 684, 77  (2010);

S. H. Hendi, S. Panahiyan and E. Mahmoudi, Eur. Phys. J. C 74,
3079 (2014).

\bibitem{Brewin} L. Brewin, Gen. Relativ. Gravit. 39, 521 (2007).

\bibitem{Comer} G. L. Comer, Class. Quantum Gravit. 9, 947 (1992).


\bibitem{Usual} D. Chaudhuri, Phys. Rev. E 75, 021803 (2007);

D. J. Searles and D. J. Evans, J. Chem. Phys. 113, 3503 (2000).

\bibitem{GBalpha} D. Kastor, S. Ray and J. Traschen, Class. Quantum Gravit. 27, 235014
(2010);

R. G. Cai, L. M. Cao, L. Li, and R. Q. Yang, JHEP 09, 005 (2013);

S. W. Wei and Y. X. Liu, Phys. rev. D 90, 044057 (2014).

\bibitem{Quevedo1} H. Quevedo, M. N. Quevedo, A. Sanchez, and S. Taj, Phys. Scr. 89, 084007
(2014).

\bibitem{Lifshitz} J. X. Mo, X. X. Zeng, G. Q. Li, X. Jiang, W. B. Liu, JHEP 10, 056 (2013).

\end{thebibliography}
\end{document}